\documentclass[12pt]{article}
\usepackage{color}

\usepackage{amsmath}
\usepackage{amssymb}
\usepackage{graphicx}
\usepackage[margin=1in]{geometry}

\usepackage{natbib}
\bibpunct[,]{(}{)}{;}{a}{}{,}

\newcommand{\Ub}{U_\text{b}}
\newcommand{\tc}{\tau_\text{c}}
\newcommand{\phg}{p_\text{hg}}

\begin{document}

\title{The stochastic edge in adaptive evolution}

\author{\'Eric Brunet$^{*}$, Igor M. Rouzine$^{\dagger}$, and Claus O. Wilke$^{\ddagger}$}

\maketitle

\bigskip

\begin{center}
$^*$ Laboratoire de Physique Statistique, \'Ecole Normale Sup\'erieure,\\24 rue Lhomond,
75230 Paris Cedex 05, France\\
$^\dagger$ Department of Molecular Biology and Microbiology, Tufts University,\\136 Harrison Avenue, Boston, MA 02111, USA\\
$^\ddagger$ Section of Integrative Biology, Institute for Cell and Molecular Biology, and Center for Computational Biology and Bioinformatics,\\University of Texas at Austin, Austin, TX 78712, USA\\
\end{center}

\bigskip

\noindent
Running head: Stochastic edge
\bigskip


\bigskip

\noindent
Corresponding author:\\
\mbox{}~~Claus O. Wilke\\
\mbox{}~~Integrative Biology\\
\mbox{}~~\#1 University Station -- C0930\\
\mbox{}~~University of Texas, Austin, TX 78712, USA\\
\mbox{}~~cwilke@mail.utexas.edu\\
\mbox{}~~Phone: (512) 471 6028\\
\mbox{}~~Fax: (512) 471 3878

\bigskip
\newpage

\noindent
\textbf{Abstract:} In a recent article, \citet{DesaiFisher2007} proposed that the speed of adaptation in an asexual population is determined by the dynamics of the stochastic edge of the population, that is, by the emergence and subsequent establishment of rare mutants that exceed the fitness of all sequences currently present in the population. \citeauthor{DesaiFisher2007} perform an elaborate stochastic calculation of the mean time $\tau$ until a new class of mutants has been established, and interpret $1/\tau$ as the speed of adaptation. As they note, however, their calculations are valid only for moderate speeds. This limitation arises from their method to determine $\tau$: \citeauthor{DesaiFisher2007} back-extrapolate the value of $\tau$ from the best-fit class' exponential growth at infinite time. This approach is not valid when the population adapts rapidly, because in this case the best-fit class grows non-exponentially during the relevant time interval. Here, we substantially extend  \citeauthor{DesaiFisher2007}'s analysis of the stochastic edge. We show that we can apply \citeauthor{DesaiFisher2007}'s method to high speeds by either exponentially back-extrapolating from finite time or using a non-exponential back-extrapolation. Our results are compatible with predictions made using a different analytical approach \citep{Rouzineetal2003,Rouzineetal2007}, and agree well with numerical simulations.

\bigskip

\centerline{INTRODUCTION}
\bigskip

For small asexual populations and low mutation rates, the speed of adaptation is primarily limited by the availability of beneficial mutations: a mutation has the time to reach fixation before the next mutation occurs. Therefore, in this case the speed of adaptation increases linearly with population size and mutation rate. By contrast, for large asexual populations or high mutation rates, beneficial mutations are abundant. In this case, the main limit to adaptation is that many beneficial mutations are wasted: when arising on different genetic backgrounds, they cannot recombine and thus are in competition with each other. The theoretical prediction of the speed of adaptation in the latter case is a formidable challenge even for the simplest models. The earliest attempts to predict this speed go back to \citet{MaynardSmith71}, and in recent years several groups have improved upon and extended this work \citep{Barton95,Tsimringetal96,PruegelBennett97,Kessleretal97,GerrishLenski98,Orr2000,Rouzineetal2003,Rouzineetal2007,Wilke2004,DesaiFisher2007}. The recent works can be broadly subdivided into two classes: (i) so-called ``clonal-interference models'' \citep{GerrishLenski98,Orr2000,Wilke2004,ParkKrug2007}, which emphasize that different beneficial mutations have different-sized effects, and that mutations with large beneficial effects tend to outcompete mutations with small beneficial effects, and (ii) models in which all mutations have the same effect $s$ \citep{Tsimringetal96,Kessleretal97,Rouzineetal2003,Rouzineetal2007,DesaiFisher2007}. The latter type of models emphasize that in large populations, multiple beneficial mutations frequently occur in quick succession on the same genetic background. These models, however, neglect clonal-interference effects.

For the second class of models, where all mutations have the same fitness effect, each individual can be conveniently described by the number $k$ of beneficial mutations it holds. The whole adapting population can then be seen as a traveling wave \citep{Tsimringetal96,Rouzineetal2003,Rouzineetal2007} moving with time through fitness space towards increasing values of $k$. In the traveling-wave approach, the bulk of the population, for which each $k$ value is occupied by many individuals, can be accurately described using a deterministic partial differential equation. However, the partial differential equation breaks down for the rare mutants that have the highest fitness in the population, because these rare mutants are subject to substantial genetic drift and stochasticity. Therefore, the description of this stochastic edge must be approached differently, and must be coupled with the description of the bulk of the population. Specifically, the deterministic equation admits a traveling-wave solution for any velocity. The high-fitness tail of that solution ends at a finite point, which is identified with the stochastic edge. To select one solution (and thus determine the wave speed), \citet{Rouzineetal2003,Rouzineetal2007} estimated the average size of the stochastic edge using a stochastic argument, and matched this size to the solution of the deterministic equation. 
  
Recently, \citet{DesaiFisher2007} have proposed a new method to calculate the speed of adaptation for the same model. They mainly carry out an elaborate treatment of the stochastic edge, with little attention paid to the bulk of the population. The full-population model is effectively replaced with a two-class model consisting of the best-fit and the second-best-fit classes only; the best-fit class is treated stochastically, whereas the next-best class is assumed to increase exponentially in time due to selection. Beneficial mutations are neglected compared to the effect of selection, except for mutations into the best-fit class. At the very end of the derivation, the sizes of other fitness classes are estimated to provide a normalization condition.

Both \citet{Rouzineetal2003,Rouzineetal2007} and \citet{DesaiFisher2007} calculate the speed of adaptation in steady state, when mutation-selection balance maintains the shape of the traveling wave. The transient dynamics generally happen on a short timescale but are hard to quantify analytically \citep{Tsimringetal96,DesaiFisher2007}. \citet{Rouzineetal2003,Rouzineetal2007} define the speed of adaptation as the change of the population's mean number of mutations over time, $V=d\langle k\rangle/dt$.  \citet{DesaiFisher2007} consider instead the change in the population's mean fitness, $v=sV$. Both approaches consider as an intermediate quantity the lead $q$,  defined as the difference between the number of mutations of the best fit individuals and the average number of mutations in the population, and write a relation between $q$ and the mean establishment time $\tau=1/V$ of a new fitness class at the stochastic edge of the population. (Note that \citet{Rouzineetal2003,Rouzineetal2007} write the lead as $|x_0|$ rather than $q$, and derive a relation between $q$ and $V$ rather than $q$ and $\tau$. Furthermore, $k$ is in these papers the number of \emph{deleterious} rather than beneficial mutations.) We would expect both approaches to make comparable predictions for $V$, and indeed they do when the speed of adaptation is moderate. For larger speeds, we cannot compare the two approaches, because \citeauthor{DesaiFisher2007}'s derivation is valid only under the condition $V<s$ (pers. comm. from M. M. Desai). If we disregard this limitation and compare the approaches nevertheless for larger speeds, we find that  \citeauthor{DesaiFisher2007}'s $V$ deviates strongly from the one obtained by \citet{Rouzineetal2007}.

Here, our goal is to provide an extensive reanalysis of the approach of \citet{DesaiFisher2007} and to extend it to the case $V>s$. For completeness, we first rederive the relation between $q$ and $\tau$ found by \citet{DesaiFisher2007} and point out the approximations made in the process. Then, we show in two different ways how we can extend their work to larger speeds of adaptation. With our modifications, the result of \citet{DesaiFisher2007} becomes compatible with the result of  \citet{Rouzineetal2003,Rouzineetal2007}. Finally, we substantiate our claims with numerical simulations.

\bigskip

\centerline{MATERIALS AND METHODS}
\bigskip

\noindent
\textbf{Model assumptions:} We consider exactly the same model as \citet{DesaiFisher2007}. Briefly, we model a population of $N$ sequences evolving in continuous time. A sequence with $k$ beneficial mutations has fitness $sk$ (which means we assume there is no epistasis); such a sequence reproduces with rate $1+sk-\langle sk\rangle$, where $\langle sk\rangle$ is the average fitness in the population. The population size $N$ is held constant at all times by removing one random sequence from the population for every reproduction event. All sequences are equally likely to be chosen for removal, and thus have the same average death rate of 1. Mutation events are decoupled from replication events, and we assume that each sequence may independently undergo a mutation with a rate $\Ub$: if a sequence has $k$ beneficial mutations, it is removed with probability $\Ub\,dt$ and replaced by a sequence with $k+1$ beneficial mutations. Since there are no differences in mutational effects in this model, all sequences with the same number of mutations $k$ can be lumped together into one fitness class, and we refer to the number of sequences with $k$ mutations at time $t$ as $n_k(t)$.

An evolutionary model in which mutation and replication events are decoupled is called \emph{parallel mutation-selection model} \citep{Baakeetal97}. This model has a long-standing tradition in theoretical population genetics \citep{CrowKimura70}. Even though the alternative model, in which mutation and selection are coupled, may be more appropriate for rapidly evolving viral populations, both models are biologically relevant. Furthermore, in the limit of small $s$ and $\Ub$, which we consider here, the mutation and selection terms decouple, and the two models become equivalent (see e.g. \citealp{Rouzineetal2003}).

When the number $n_k(t)$ of sequences with mutation number $k$ is large enough, the evolution of $n_k(t)$ becomes nearly deterministic:
\begin{equation}
\frac{dn_k(t)}{dt}=\big(sk -\langle s k\rangle\big)n_k(t)+\Ub[n_{k-1}(t)-n_k(t)].
\label{deterministic}
\end{equation}
Sequence classes that satisfy this condition and follow Eq.~\eqref{deterministic} are called \emph{established}. However, the best-fit sequences in the population are not numerous enough for a deterministic description, and stochasticity and genetic drift play an important role in their evolution. We make the approximation that we give the best-fit fitness class (the class corresponding to the largest $k$ with $n_k(t)>0$) a precise stochastic treatment, while we regard all other classes as established and treat them deterministically. The validity of this approximation has been discussed in detail \citep{Rouzineetal2003,DesaiFisher2007,Rouzineetal2007}; in particular, it was shown by \citet{Rouzineetal2007} that this approximation is valid if the speed of adaptation is much larger than $\Ub$. Moreover, we check this approximation numerically in the present work. We shall refer to the one stochastic class as the \emph{stochastic edge}, and denote the value of $k$ for that class by $k_0$.

Let $\langle k\rangle$ be the mean number of mutations in the population. We define the \emph{lead} $q$ as $q=k_0-\langle k\rangle$. The lead is the distance from the stochastic edge to the population center. By the definition of fitness in the model, sequences at the stochastic edge have a fitness advantage of $sq$ over the bulk of the population, and sequences in the first established (i.e., second-best) class have a fitness advantage of $s(q-1)$. Following \citet{DesaiFisher2007}, we make the approximation that the second-best class behaves deterministically according to Eq.~\eqref{deterministic}, and, neglecting incoming mutations from the third best class and outgoing mutations to the best class, that it grows approximately exponentially with rate $s(q-1)$. (We shall discuss or check numericaly the validity of these approximations later on.) While the second-best class is growing, any beneficial mutations that occur to sequences in this class feed the best class. Even though any individual mutant that arrives in the best class has a substantial probability of being lost to drift, the ongoing feeding of the best class guarantees that this class itself will become established at some point in time. At this point, the newly-established fitness class becomes the second-best class (which, as we assumed, grows deterministically), a new stochastic edge develops at $k_0+1$, and the process repeats.

Note that during one cycle, the values of $\langle k\rangle$ and $q$ change smoothly by one unit, but we ignore that change and assume that $q$ remains constant from the creation of a new best class to its establishment. Therefore, the whole approach is only valid if $q$ is large enough so that it makes sense to neglect a change of order 1 in $q$. We assume also that the stochastic edge becomes established when its size gets large enough compared to $1/(sq)$, which is a well known stochastic threshold \citep{MaynardSmith71,Barton95,Rouzineetal2001}. This assumption makes sense only if $sq\ll1$. Finally, we assume that the stochastic edge does not produce any mutant until it is established, which implies $\Ub\ll sq$. These conditions imply, of course, $s\ll1$. Note that $q$ is not a parameter of the model, but a derived quantity. Therefore, all these assumptions must be checked \textit{a posteriori} once $q$ is computed as a function of the parameters $N$, $s$, and $\Ub$.

\noindent
\textbf{Simulations:}
We carried out three types of numeric simulations: fully stochastic whole-population simulations, semideterministic whole-population simulations, and stochastic-edge simulations. The first two ones are simulations of the whole population, whereas the third one is a simulation of the growth of the best-fit class only assuming it is fed by an exponentially growing second-best-fit class. Details are given below. In all cases, we simulated continuous time by subdividing one generation into small time steps of length $\delta t$, and updated the simulation after every such time step. In all results reported, $\delta t$ was at most 0.01.

We used both the GNU Scientific Library \citep{GSL} and the library libRmath from the R project \citep{Rproject} for generation of Poisson, multinomially, and hypergeometrically distributed random numbers. Source code to all simulations is available upon request from C.O.W.

\noindent
\textbf{Fully stochastic whole-population simulations:} For each fitness class $k$, we kept track of a random variable $n_k(t)$ representing the class size at time $t$. In each time step, we first calculated the number of offspring $o_k$ in fitness class $k$. The $o_k$ are Poisson random variables with mean $n_k(t)(1+sk-\langle sk\rangle) \delta t$. We then calculated the number of deaths $d_k$ in each class. The total number of deaths $D=\sum_k d_k$ in one time step equals the number of new offspring in that time step, $D=\sum_k o_k$. We generated the $d_k$'s by drawing a single set of multinomially distributed random numbers with means $\langle d_k\rangle=Dn_k(t)/N$ and $\sum_k d_k=D$. If we obtained one or more $d_k$ with $d_k>n_k(t)$, we redrew the entire set of $d_k$'s. We then computed the state of the population after selection but before mutation as $n'_k=n_k(t)+o_k-d_k$. Next, we generated mutations. For each class $k$, we generated a binomially distributed random variable $m_k$ with mean $\langle m_k\rangle = n'_k\Ub\delta t$ and $n'_k$ trials. We then updated the population to $n_k(t+\delta t)=n'_k+m_{k-1}-m_k$.

The usage of the multinomial distribution to generate $d_k$'s is an approximation, as the distribution of the $d_k$'s is actually hypergeometric. [The hypergeometric distribution describes the probability $\phg(d,D;n,N-n)$ to obtain $d$ white balls after $D$ random draws from an urn containing $n$ white and $N-n$ black balls, and is given by $\phg(d,D;n,N-n)=\binom{n}{d}\binom{N-n}{D-d}/\binom{N}{D}$.] We also implemented hypergeometric sampling of deaths, by generating the random variables $d_k$ one by one, going from the best-fit class to the worst-fit class with the probabilities $\text{Prob}(d_k)=\phg[d_k, D-\sum_{i>k}d_i, n_k(t), \sum_{i<k}n_i(t)]$. We found that the generation of hypergeometrically distributed random variables was much slower than multinomial sampling (up to a factor of 1000) and caused numeric instabilities at large $N\gtrsim 10^8$, even when using an efficient numerical algorithm \citep{KachitvichyanukulSchmeiser85}. For $N<10^8$, simulation results with multinomial sampling of deaths and hypergeometric sampling of deaths were virtually identical.

We measure the speed of adaptation $V$ in steady state, when the population can be considered a traveling wave \citep{Rouzineetal2003,Rouzineetal2007}. We know of no good theory for predicting how long it takes for the population to reach steady state, but simulations indicate that equilibration proceeds rapidly (see also \citealt{Tsimringetal96}). In our simulations, we considered the population as equilibrated when at least 10 new fitness classes had been established. We then measured the time $\Delta t$ it took the population to establish 40 additional fitness classes, and calculated $V$ as $40/\Delta t$. We averaged $V$ over 10 independent replicates.

\noindent
\textbf{Semideterministic simulations:} For each fitness class $k$, we kept track of a variable $n_k(t)$ representing the class size at time $t$. We updated the size of the stochastic edge class $n_{k_0}(t)$ stochastically, and all other variables $n_k(t)$ deterministically. As in the case of the fully stochastic simulations, in each time step we first calculated the number of offspring $o_k$ in fitness class $k$. For $k<k_0$, $o_k=n_k(t)(1+sk-\langle sk\rangle) \delta t$. At the stochastic edge, $o_{k_0}$ is a Poisson-distributed random variable with mean $n_{k_0}(t)(1+sk-\langle sk\rangle) \delta t$. We then calculated the number of deaths $d_k$. The total number of deaths required is $D=\sum_k o_k$. At the stochastic edge, $d_{k_0}$ is a Poisson-distributed random variable with mean $Dn_{k_0}(t)/N$. We set $d_{k_0}=n_{k_0}$ if $d_{k_0}>n_{k_0}$. For $k<k_0$, we calculated $d_k=(D-d_{k_0})n_k(t)/(N-n_{k_0})$. We then computed the state of the population after selection but before mutation as $n'_k=n_k(t)+o_k-d_k$. Next, we generated mutations. At the stochastic edge, $m_{k_0}=0$ (the stochastic edge does not produce beneficial mutations). For the second-best class, $m_{k_0-1}$ is a Poisson-distributed random variable with mean $n'_{k_0-1}\Ub\delta t$. For all other $k<k_0-1$, $m_k=n'_k\Ub\delta t$. We then updated the population to $n_k(t+\delta t)=n'_k+m_{k-1}-m_k$. [In theory, this procedure can lead to a negative $n_{k_0-1}(t+\delta t)$. However, this extremely unlikely event never actually occurred in our simulations.]  Finally, if $n_{k_0}(t)>1/(sq)$, we designated the current stochastic edge class as established, and set $k_0$ to $k_0+1$.

We measured the speed of adaptation as in the fully stochastic full-population simulations.

\noindent
\textbf{Stochastic-edge simulations:} We kept track of a single random variable $n(t)$ representing the best-fit class in the population (the stochastic edge), which was set to zero at $t=0$. Assuming that the population of the second-best-fit class was $e^{s(q-1)t}/(sq)$, we generated at each time step three Poisson random variables $o$, $d$, and $m$, representing the number of offspring, deaths, and incoming mutations in the best-fit class, with means $(1+sq)n(t)\delta t$, $n(t)\delta t$, and $\Ub e^{s(q-1)t}\delta t/(sq)$, respectively, and updated $n(t)$ as $n(t+\delta t)=n(t) + o-d+m$. All measures reported in the Results section were obtained by averaging over 500 independent realizations of the simulation.

\bigskip

\centerline{RESULTS AND DISCUSSION}
\bigskip

\noindent
\textbf{Rederivation of the Desai-Fisher results:} In this section, we rederive the main results of \citet{DesaiFisher2007}, using methods very similar to theirs, but with some simplifications. This section does not contain any new results; it is included here because we need to point out the various approximations made by \citet{DesaiFisher2007} before discussing them, and also because we believe that an alternative presentation of their non-trivial results may be helpful to many readers.

\citet{DesaiFisher2007} define the \emph{establishment time} $\tau$ as the time from the establishment of one new fitness class to the establishment of the next better fitness class. Their approach is based on an elaborate probabilistic calculation of the establishment time $\tau$ of a fitness class with advantage $sq$, given that this class is fed beneficial mutations from the exponentially-growing second-best class. Since for large $N$ every beneficial mutation that arrives in the best-fit class forms a clone that is independent of all other clones in the best-fit class, the growth (and potential establishment or extinction) of a single clone can be described using continuous-time branching theory. The treatment of a single clone is standard \citep{AthreyaNey72}; the single clone follows a birth-and-death process with birth rate $1+sq$ and death rate 1. The probability-generating function for the size $m(t)$ of a clone at time $t$ that had size 1 at time $t=0$ is given by \citep{AthreyaNey72}
\begin{equation}\label{eq:single-clone-gen}
  G(z,t) = \langle z^{m(t)}\rangle = \frac{(z-1)(1-e^{sqt})+zsq}{(z-1)[1-(1+sq)e^{sqt}]+zsq}.
\end{equation}
Given this result for a stochastically growing individual clone, we now wish to study the size $n(t)$ of the best-fit class at time $t$. This class grows by itself with a rate $sq$ and is fed by the next-best class, which grows at a rate $s(q-1)$. We call $f(t)$ the size of the next-best class, and we shall assume later on that
\begin{equation}
  f(t) = \frac{1}{sq} e^{s(q-1)t}.
\label{secondbestfitclass}
\end{equation}
Note that we assume a deterministic growth for this next-best class, and we neglect changes in $f(t)$ due to both outflow of beneficial mutations from the second-best to the best-fit class and inflow of beneficial mutations from the third-best to the second-best class. We set the origin of time such that $t=0$ when $f(t)=1/(sq)$, which is the well-known stochastic threshold for a clone with fitness advantage $sq$ \citep{MaynardSmith71,Barton95,Rouzineetal2001}: a clone whose size far exceeds this threshold grows essentially deterministically, whereas a clone whose size falls far below this threshold is subject to genetic drift. The idea is that this second-best-fit class just got established at time $t=0$ and was the previous stochastic best-fit class at times $t<0$.

As the size $n(t)$ of the best-fit class grows with rate $sq$, \citet{DesaiFisher2007} suggest to write for large $t$
\begin{equation}\label{eq:deftau}
  n(t) = \frac{1}{sq}e^{sq(t-\tau)},
\end{equation}
where $\tau$ is some random variable. Intuitively, we might interpret $\tau$ as the time at which the new best-fit class \emph{appears} to have reached the stochastic threshold, when sampled at later times when it is already deterministic. Of course, $n(t)$ is really a random variable which in the $t\to\infty$ limit converges to an exponential growth, and $n(\tau)$ has no reason to be equal to $1/(sq)$. But if $n(t)$ had a deterministic exponential growth at all time, then $\tau$ would be the time for which $n(t)$ had reached $1/(sq)$. In this case, it would take a (random) time $\tau$ to build a new established class from a class that had just crossed the stochastic threshold, the population would move by one fitness class during a time interval $\tau$, and the speed of adaptation would be simply $1/\langle\tau\rangle$, where $\langle\tau\rangle$ is the average of $\tau$. However, things are more complicated than this intuitive picture suggests. In Eq.~\eqref{eq:deftau}, the random variable $\tau$ has a distribution which depends on the time $t$ at which n(t) is measured and we shall from now on write $\tau(t)$ rather than simply $\tau$. When evaluating the speed of adaptation $1/\langle\tau(t)\rangle$, we have to choose a time $t$ at which the average is taken.  \citet{DesaiFisher2007} chose to take $t=\infty$ and, therefore, to define the establishment time as $\langle\tau(\infty)\rangle$. This choice of $t$ is however arbitrary and we shall argue later on that it makes more sense to choose $t$ on the order of $\langle\tau\rangle$. As we shall see, when $q$ is large $\langle\tau(t)\rangle$ converges quite slowly to $\langle\tau(\infty)\rangle$, and the two expressions give different results. For this reason, \citet{DesaiFisher2007} considered only moderately large $q$, for which $\langle\tau(\infty)\rangle$ is a good approximation of $\langle\tau(t)\rangle$.

Note that replacing the threshold $1/(sq)$ in Eqs.~\eqref{secondbestfitclass} and \eqref{eq:deftau} by $\alpha/(sq)$ results only in an additional factor $\alpha^{-1/q}$ inside the large logarithm in the final result for $\langle\tau(\infty)\rangle$ [see Eq.~\eqref{eq:tau-ave1} below]. Because $q$ is assumed to be large, the effect of that change is minor.

We put aside for the moment the problem of choosing the best value of $t$ in $\langle\tau(t)\rangle$  and focus on calculating the cumulants of $\tau(t)$. We first calculate the probability-generating function $\langle z^{n(t)}\rangle$ of $n(t)$, under the assumption that the best-fit class is fed by mutations from the second-best class, which itself has size $f(t)$ at time $t$. In this calculation, we neglect beneficial mutations produced by the best-fit class, as mutations are rare and the number of sequences in this class is small. Assuming that the process starts at some time $T_0$ with $n(T_0)=0$, we write $n(t)=\sum_{T_0<t'<t} m_{t'}(t)$, where $m_{t'}(t)$ is the contribution at time $t$ of a new clone if such a clone appeared at time $t'$. With a probability $f(t')\Ub dt'$, a clone actually appeared at time $t'$ and, according to Eq.~\eqref{eq:single-clone-gen}, we have $\langle z^{m_{t'}(t)} \rangle=G(z,t-t')$. With a probability $1-f(t')\Ub dt'$, no clone appeared at time $t'$, we have $m_{t'}(t)=0$ and, of course, $\langle z^{m_{t'}(t)} \rangle=1$. As all the $m_{t'}(t)$ for a given $t$ are independent random numbers, we can write $z^{n(t)}=\prod_{T_0<t'<t} z^{m_{t'}(t)}$ and average independently all the terms in the product. We obtain
\begin{align}
\langle z^{n(t)}\rangle &= \prod_{T_0< t' < t}\Big(f(t')\Ub dt' G(z,t-t') + [1-f(t')\Ub dt']\Big) \notag\\
 &= \exp\Big[ \sum_{T_0<t'<t} \ln\Big(1 + dt'\Ub f(t')[ G(z,t-t') - 1 ]\Big)\Big].
\end{align}
As $dt'$ is infinitely small, we have $\ln(1 + dt'C)=dt'C$, and we recognize that the summation is actualy an integral. Therefore,
\begin{align}
\langle z^{n(t)}\rangle &= \exp\Big( \Ub \int_{T_0}^t dt' f(t')[ G(z,t-t') - 1 ]\Big)
\label{eq:DFEq25x}
 \\
  &=\exp\Big( \Ub \int_0^{t-T_0}dt' f(t-t')[ G(z,t') - 1 ]\Big).
\label{eq:DFEq25}
\end{align}
This equation with $T_0=-\infty$ corresponds to Eq.~(24) of \citet{DesaiFisher2007}.

In Eq.~\eqref{eq:DFEq25}, we use the $G(z,t)$ given in Eq.~\eqref{eq:single-clone-gen} and the $f(t)$ given in Eq.~\eqref{secondbestfitclass}, change the variable of integration to $v=(1+sq)(1-z)e^{sqt'}/[zsq-(1-z)]$, and obtain
\begin{equation}\label{eq:DFEq28}
  \langle z^{n(t)}\rangle = \exp\Bigg[-\frac{\Ub\Big[\Big(\frac{zsq}{1-z}-1\Big)e^{-sqt}\Big]^{(1/q)-1}}{sq(1+sq)^{1/q}}
          \times\int_{\frac{(1-z)(1+sq)}{zsq-(1-z)}}^{\frac{(1-z)(1+sq)}{zsq-(1-z)}e^{sq(t-T_0)}}
			\frac{v^{(1/q)-1}dv}{v+1}\Bigg].
\end{equation}
[This equation corresponds to Eq.~(27) of \citet{DesaiFisher2007}. Note that because of the way the change of variable was done, it is only correct for $zsq>1-z$.] To compute the cumulants of $\tau(t)$, it is easier to rewrite Eq.~\eqref{eq:deftau} as
\begin{equation}\label{defofx}
 n(t)={1\over sq} x(t) e^{sqt}
\end{equation}
with the random variable $x(t) = e^{-sq\tau (t)}$; the cumulants of $\ln x(t)$ differ from the cumulants of $\tau(t)$ only by a constant multiplicative factor. We obtain the generating function of $x(t)$ from Eq.~\eqref{eq:DFEq28} using $\langle e^{-\lambda x(t)}\rangle=\langle z^{n(t)}\rangle$ for $z=\exp\big(-\lambda sq e^{-sqt}\big)$. For now, we are only interested in the limit of infinite time. Making the substitution for $z$ and taking the limit $t\to\infty$ while holding $\lambda$ constant, we find
\begin{equation}\label{eq:mgf-x-integral}
  \langle e^{-\lambda x(\infty)}\rangle = \exp\Big[-\frac{\Ub}{sq} \frac{\lambda^{1-1/q}}{(1+sq)^{1/q}}
       \times\int_0^{\lambda (1+sq) e^{-sqT_0}} \frac{v^{(1/q)-1}dv}{v+1}\Big].
\end{equation}
In this expression, $T_0$ is the starting time at which the second-best class begins feeding the best-fit class. The second-best class can start producing mutants when its size is of order 1, which happens at large negative times. Unfortunately, Eq.~\eqref{eq:mgf-x-integral} is, strictly speaking, not valid if $T_0<0$, as we obtained it by using, in Eq.~\eqref{eq:DFEq25x}, the expression Eq.~\eqref{secondbestfitclass} for the size of the second-best fit class, which is correct only for $t>0$. However, as we assumed $\Ub /(sq)\ll1$, the mutation events from the second-best class at any negative time are very rare, so that we may expect that the final result will be dominated only by the events with $t>0$  and that it will not depend much on the value of $T_0$, as long as $T_0$ is a negative number. One way to check the validity of this assumption is to verify that we reach the same results for $T_0=-\infty$ (equivalent to the assumption that Eq.~\eqref{secondbestfitclass} is a good approximation for the size of the second-best class at negative times) and for $T_0=0$ (equivalent to the assumption that the second-best class is empty at negative times). Therefore, we first follow \citet{DesaiFisher2007}  by taking the limit $T_0\to-\infty$, and, at the end of this section, we will consider briefly the case $T_0=0$ to validate this approximation. For $T_0=-\infty$, using $\int_0^\infty dv\, v^{(1/q)-1}/(v+1)=\pi/\sin(\pi/q)$, we obtain
\begin{equation}\label{eq:mgf-x}
  \langle e^{-\lambda x(\infty)}\rangle = \exp(-b\lambda^{1-1/q})
\end{equation}
with
\begin{equation}\label{defb}
  b = \frac{\pi \Ub}{sq(1+sq)^{1/q}\sin(\pi/q)}\approx
  \frac{\pi \Ub}{sq\sin(\pi/q)}.
\end{equation}
[The first expression for $b$ in Eq.~\eqref{defb} is exact, but we will only use the second, approximate expression in the following of the paper as we need $s\ll1$ anyway in the biological applications of that model. In fact, we will often use $b\approx\Ub/s$ when we suppose $q\gg1$.]

Eq.~\eqref{eq:mgf-x} is the generating function of $x(\infty)$, but we need the generating function of $\ln x(\infty)$. For any random variable $x$, we can turn the former into the latter using the following identity, which is valid for $\mu<0$ and follows from the definition of the Gamma function:
\begin{equation}\label{eq:xpowmu}
  \langle x^\mu \rangle  = \frac{1}{\Gamma(-\mu)} \int_0^\infty d\lambda\, \lambda^{-\mu-1} \langle e^{-\lambda x}\rangle.
\end{equation}
(Actually, the equality holds without the averages.) 
Then, expanding $\ln \langle x^\mu\rangle$ in powers of $\mu$ allows us to recover all the cumulants of $\ln x$:
\begin{equation}\label{eq:xcumu}
 \ln \langle x^\mu\rangle = \mu \langle \ln x\rangle + \frac{\mu^2}{2}\text{Var}[\ln x] + {\cal O}(\mu^3).
\end{equation}
Alternatively, if all we need is $\langle\ln x\rangle$, we can integrate by part $\lambda^{-\mu-1}$ in Eq.~\eqref{eq:xpowmu} (assuming that $\langle e^{-\lambda x}\rangle$ goes to 0 for large $\lambda$) and expand directly to the first order in $\mu$. We obtain
\begin{equation}\label{eq:xcumubypart}
\langle\ln x\rangle = -\gamma + \int_0^\infty d\lambda\, \ln(\lambda) {d\over d\lambda}\langle e^{-\lambda x}\rangle,
\end{equation}
where $\gamma=-\Gamma'(1)\approx0.5772$ is the Euler gamma constant.
Applying this procedure to the random variable $x(\infty)$, we get from Eq.~\eqref{eq:xpowmu}
\begin{align}\label{eq:avexpowmu}
   \langle x(\infty)^\mu\rangle &= \frac{1}{\Gamma(-\mu)} \int_0^\infty d\lambda\, \lambda^{-\mu-1} \exp(-b\lambda^{1-1/q})
= \frac{\Gamma\big(1-\frac{\mu q}{q-1}\big)}{\Gamma(1-\mu)}
   b^{\frac{\mu q}{q-1}}.
\end{align}
Making use of the expansion $\ln \Gamma(1-\epsilon)=\gamma \epsilon + (\pi\epsilon)^2/12 + {\cal O}(\epsilon^3)$, we obtain from Eq.~\eqref{eq:xcumu}
\begin{align}\label{eq:lnx-ave1}
  \langle \ln x(\infty)\rangle = \frac{q}{q-1}\ln(be^{\gamma/q}),\\
  \text{Var}[\ln x(\infty)] = \frac{\pi^2}{6}\Big[\Big(\frac{q}{q-1}\Big)^2-1\Big].
\end{align}
Converting $\ln x(\infty)$ back into $\tau(\infty)$, we arrive at our final expressions
\begin{align}\label{eq:tau-ave1}
    \langle \tau{(\infty)}\rangle &= \frac{1}{s(q-1)}\ln\Big(\frac{1}{be^{\gamma/q}}\Big)\notag\\
	& \approx \frac{1}{s(q-1)}\ln\Big(\frac{sq\sin(\pi/q)}{\Ub\pi e^{\gamma/q}}\Big)
\end{align}
and
\begin{equation}\label{eq:Var-of-tau}
  \text{Var}[\tau{(\infty)}] = \frac{\pi^2}{6}\Big[\frac{1}{[s(q-1)]^2}-\frac{1}{(sq)^2}\Big].
\end{equation}
We emphasize that these quantities were obtained in the limit $t\to\infty$.

When we compare our results for mean and variance of $\tau{(\infty)}$ to the results of \citet{DesaiFisher2007}, we find that our expression for the variance agrees with their Eq.~(37). Our expression for $\langle \tau({\infty})\rangle$ is similar to their Eq.~(36), except that the factor $q$ in the logarithm was accidently replaced by a factor $q-1$ in \citet{DesaiFisher2007} (Michael Desai, pers.\ communication). As \citeauthor{DesaiFisher2007}, we neglected the factor $(1+sq)^{1/q}$ in the expression~\eqref{defb} of $b$ as we need $s\ll1$ in the context of the full biological model.

We now consider what happens if we use $T_0=0$ (and $sq\ll1$) in Eq.~\eqref{eq:mgf-x-integral} instead of $T_0=-\infty$. Clearly, for large $q$, this integral is dominated by small $v$, so the value of the upper bound should not matter much to the final result. Indeed, if $\lambda$ is not too small, we have:
\begin{align}\label{simpT0}
\int_0^{\lambda  } \frac{v^{(1/q)-1}dv}{v+1}
&=\int_0^{\infty} \frac{v^{(1/q)-1}dv}{v+1}
-\int_{\lambda  }^{\infty} \frac{v^{(1/q)-1}dv}{v+1}\notag\\
&\approx {\pi\over\sin(\pi/q)} -\ln\left(1+{1\over \lambda }\right)
\approx {\pi\over\sin(\pi/q)}.
\end{align}
We neglected $v^{1/q}$ in the last integral, which is valid if $\lambda$ is large enough, namely if either $\lambda>1$ or $-\ln\lambda \ll q$. The same condition on $\lambda$ allows the last simplification in Eq.~\eqref{simpT0}. Therefore, the generating function Eq.~\eqref{eq:mgf-x-integral} is identical for $T_0=0$ or $T_0=-\infty$, except for very small $\lambda$, and the probability distribution function of $x(\infty)$ does not depend on $T_0$ except for very large values of $x(\infty)$ such that $\ln x(\infty)\gg q$. As it is easy to check from Eq.~\eqref{eq:mgf-x} that Eq.~\eqref{eq:xcumubypart} is dominated by values of $\lambda$ of order $1/b\approx s/\Ub$, we finally obtain that the result for $\langle\ln x(\infty)\rangle$ and hence $\langle\tau(\infty)\rangle$ is approximatively the same for $T_0=0$ or $T_0=-\infty$ if either $s/\Ub>1$ or $q\gg\ln(\Ub/s)$. As \citeauthor{DesaiFisher2007} assumed $s/\Ub\gg1$ in their work, their approximation of taking $T_0=-\infty$ is justified.

As a side matter, note that the generating function Eq.~\eqref{eq:mgf-x} describes a distribution with a long tail; in particular, the average of $x(\infty)$ is infinite, which is not biologically possible and is an artefact of taking $T_0=-\infty$. If we were interested in the average of $x(\infty)$, we would need to keep $T_0$ finite and we would obtain, after some algebra, $\langle x(\infty)\rangle=(\Ub/s)e^{-sT_0}$. 
\bigskip

\noindent
\textbf{The case of large $q$:} In general, a weak selective pressure ($s\ll 1$) results in a broad fitness distribution, $q\gg 1$. In order to gain better insight into the predictions of Eq.~\eqref{eq:tau-ave1} for this case, we consider the limit $q$ large and $s$ small. We find
\begin{equation}
  \label{tauinfty}
  \langle \tau{(\infty)}\rangle \approx \frac{1}{sq}\ln(s/\Ub).
\end{equation}
By studying the deterministic evolution of the bulk of the population in the same $q\gg1$ limit, \citet{Rouzineetal2007} obtained in their Eq.~(39) a relation very similar to Eq.~\eqref{tauinfty}. Using $\tau$ and $q$ instead of the notations $V=1/\tau$ and $x_0=-q$ of the cited work, we can write their result as
\begin{align}
\tau&={1\over sq}\left[\ln\big(1/ (\Ub\tau)\big)-1\right]= {1\over sq}\Big[\ln\Big(\frac{sq}{\Ub|\ln(e\Ub\tau)|}\Big)-1\Big]\notag\\
     &={1\over sq}\big[\ln(sq/\Ub)-1-\ln |\ln(e\Ub\tau)|\big].
\label{eq:rouzine}
\end{align}
Ignoring subleading corrections, we find that the main difference between  Eq.~\eqref{tauinfty} and Eq.~\eqref{eq:rouzine} is a term $q$ within the logarithm, which can become large in some situations.

We claim that when $q$ is large, Eq.~\eqref{tauinfty} is not an accurate prediction for the mean establishment time. In particular, we obtain $\langle\tau(\infty)\rangle<0$ for $s<\Ub$. (Note that we assume throughout this work that $sq\gg \Ub$, but unlike \citet{DesaiFisher2007}, we do not require $s>\Ub$.) This result is problematic, because the whole point of this calculation was to interpret $\langle\tau(\infty)\rangle$ as the mean time between the establishment of a best-fit class and the establishment of the next best-fit class in the full model describing a population of $N$ sequences. Clearly, the establishment time in the full model cannot be negative, and this result would seem to suggest that the whole approach of approximating the full model by the sole behaviour of its stochastic edge does not work for large values of $q$. However, we believe that the method can be fixed by replacing some of the assumptions that led to Eq.~\eqref{eq:tau-ave1} by improved and more accurate assumptions.

\bigbreak \noindent \textbf{Approximations made in Desai and Fisher's approach: } 
\citet{DesaiFisher2007} made several approximations in order to obtain the relation Eq.~\eqref{eq:tau-ave1} between the establishment time and the lead $q$:
\begin{enumerate}
\item\label{ap:det} All the classes are evolving deterministically, except the stochastic edge.
\item\label{ap:lead} The lead $q$ does not vary in time between the creation and the establishment of a new mutant class.
\item\label{ap:se} The stochastic edge does not produce any mutant until it is established.
\item\label{ap:exp} The second-best-fit class has an exactly exponential growth with a rate $s(q-1)$, as in Eq.~\eqref{secondbestfitclass}.
\item\label{ap:T0} One can take the limit $T_0\to-\infty$ when evaluating the mean establishment time.
\item\label{ap:fit} At large times, the size $n(t)$ of the stochastic edge is well fit by an exponential growth of rate $sq$, as in Eq.~\eqref{eq:deftau} or Eq.~\eqref{defofx}.
\item\label{ap:t} One can interpret the establishment time as $\langle\tau(t)\rangle$ for $t\to\infty$.
\end{enumerate}
\citet{DesaiFisher2007} discussed the validity of these approximations in the context of their parameter range of interest, i.e. for moderate $q$ (see their Appendices E through G). We reevaluate the approximations here in the context of large $q$.

\citet{Rouzineetal2003,Rouzineetal2007} gave detailed analytical arguments why Approximation~\ref{ap:det} is valid for $q\gg \Ub/s$. In the present work, we verify this approximation numerically, using the semideterministic full-population simulation. Getting rid of this approximation and treating all classes stochastically is a formidable mathematical challenge which would be of limited interest because the approximation is quite good.

Approximations~\ref{ap:lead} and~\ref{ap:se} are valid in, respectively, the limits $q\gg1$ and $sq\gg U_b$, which we have assumed throughout. For more moderate values of $q$ (between 2 and 5), \citet{DesaiFisher2007} discussed the validity of Approximation~\ref{ap:lead} in their Appendix~H.

Approximation~\ref{ap:exp} is more problematic. Saying that the second-best-fit class grows exponentially implies that we are ignoring the contribution from mutations originating in the third-best-fit class. On one hand the mutation rate $\Ub$ is supposed to be small compared to the effect of selection $s(q-1)$, but on the other hand the third-best-fit class is much larger than the second-best class. In Appendix~A, we present an argument indicating that Approximation~\ref{ap:exp} is justified only at smaller times, and is incorrect by a large factor for values of $t$ close to the establishment time, which is unfortunately precisely the time at which most of the mutations occur. To what extent this deviation from Approximation~\ref{ap:exp} affects our final result is difficult to assess at this point. Improving upon this approximation would require having a theory of at least the third-best-fit class.

We have already discussed the validity of Approximation~\ref{ap:T0}. 

Approximations~\ref{ap:fit} and~\ref{ap:t} are closely related: one can always decide to write Eq.~\eqref{eq:deftau} for a well chosen time-dependent random variable $\tau(t)$. But saying that $n(t)$ is well fit by an exponential (Approximation~\ref{ap:fit}) is then equivalent to saying that $\tau(t)$ actually does not depend too much on time and that, consequently, one can choose any value of $t$ to evaluate the establishment time, including $t=\infty$ (Approximation~\ref{ap:t}). But, as we shall now argue, $n(t)$ is not well fit by an exponential growth for large $q$. This implies that $\tau(t)$ has a strong $t$ dependence and that choosing the best value of $t$ when evaluating the establishment time is important; we shall argue that the proper value of $t$ is of the order of the establishment time. Alternatively, one can get rid of Approximation~\ref{ap:fit} and replace Eq.~\eqref{eq:deftau} by a better fit of $n(t)$. When carrying out this procedure, we find that the new random variable $\tau$ has indeed a weak time dependence and taking the limit $t\to\infty$ makes sense. We shall presently explore both possible improvements.
\bigskip

\noindent
\textbf{Finite extrapolation time:}
We are still fitting the best-fit class by an exponential, as in Eq.~\eqref{eq:deftau} or Eq.~\eqref{defofx}, but this time we try to evaluate $\langle\tau(t)\rangle$ for some finite time $t$.
We go back to Eq.~\eqref{eq:DFEq28}, set $T_0=-\infty$, and substitute $z=\exp(-\lambda sqe^{-sqt})$ as before. However, this time we keep all terms to the first order in $e^{-sqt}$. We find
\begin{equation}\label{eq:momentx1}
 \langle e^{-\lambda x(t)}\rangle = \exp\Bigg[-\frac{\Ub\lambda^{1-1/q}}{sq(1+sq)^{1/q}}
         \Big[1+\lambda\frac{q-1}{q}\Big(\frac{sq}{2}+1\Big)e^{-sqt}\Big]
          \int_{(1+sq)\lambda e^{-sqt}}^\infty \frac{v^{1/q-1}dv}{v+1}\Bigg].
\end{equation}
For $\lambda e^{-sqt}\ll1$, the integral assumes the value
\begin{equation}\label{integ}
  \frac{\pi}{\sin(\pi/q)} - q(1+sq)^{1/q}\lambda^{1/q}e^{-st} + {\cal O}(e^{-s(q+1)t}).
\end{equation}
Note that the small term appearing in the integral of Eq.~\eqref{eq:momentx1} is $e^{-sqt}$, but the first order correction for finite time is actually proportional to $e^{-st}$, which is much larger. Compared to this correction, we neglect the term proportional to $e^{-sqt}$ before the integral in Eq.~\eqref{eq:momentx1}, and find
\begin{equation}\label{eq:momentxfinal}
 \langle e^{-\lambda x(t)}\rangle \approx \exp\Big[-b\lambda^{1-1/q}+\frac{\Ub}{s} \lambda e^{-st}\Big]\qquad\text{for $\lambda e^{-sqt}\ll1$}.
\end{equation}
[\citet{DesaiFisher2007} write a similar expression in their Eq.~(G2), but do not exploit it.]
When $\lambda e^{-sqt}$ is not small but $q$ is large, we can obtain another expression by neglecting the $v^{(1/q)}$ in Eq.~\eqref{eq:DFEq28}.
Assuming $sq$ small, we obtain after some algebra
\begin{equation}\label{eq:momentxfinal2}
\langle e^{-\lambda x(t)}\rangle \approx \exp\Big[-\frac{\Ub}{sq}e^{-s(q-1)t}\times{\lambda e^{-sqt}\over \lambda e^{-sqt}-1}\ln\big(\lambda e^{-sqt}\big)\Big]\quad\text{for $e^{-q}\ll\lambda e^{-sqt}\ll1/(sq)$}.
\end{equation}
Note that Eq.~\eqref{eq:momentxfinal} and Eq.~\eqref{eq:momentxfinal2} are both valid in the range $e^{-q}\ll\lambda e^{-sqt}\ll1$.

We want, as before, to compute $\langle x(t)^{\mu}\rangle$ by using
Eq.~\eqref{eq:momentxfinal} into Eq.~\eqref{eq:xpowmu}. Expanding inside the integral in powers of the small parameter $\exp(-st)$, we would get:
\begin{equation}\label{eq:infinite}
\langle x(t)^\mu\rangle \approx {1\over\Gamma(-\mu)}\sum_{n\ge0} {1\over n!}\left( \Ub\over s\right)^n e^{-nst}\int_0^{\infty} d\lambda\, \lambda^{-\mu-1} \exp\Big[-b\lambda^{1-1/q}\Big]\lambda^n,
\end{equation}
but writing this equation is not justified \textit{a priori}, because we may not use Eq.~\eqref{eq:momentxfinal} for arbitrarily large $\lambda$, and Eq.~\eqref{eq:infinite} is actually a divergent series. We will show, however, that the first terms of that series are nevertheless correct. Indeed, the integral in the $n$-th order term of the series Eq.~\eqref{eq:infinite} is mainly contributed from values of $\lambda$ of order $n/b\approx ns/\Ub$. (This result is obtained by looking at the maximum of the integrand, in the limit of large $q$ and large $n$.) Given the validity range of Eq.~\eqref{eq:momentxfinal}, this means that the series Eq.~\eqref{eq:momentxfinal} is correct up to $n=n_\text{max}$ with $n_\text{max}\ll (\Ub/s) e^{sqt}$. With this in mind, we compute the integrals and find
\begin{equation}
  \langle x(t)^\mu\rangle \approx\frac{\Gamma\Big(1-\frac{\mu q}{q-1}\Big)}{\Gamma(1-\mu)} b^{\frac{\mu q}{q-1}} 
   \Bigg[ 1-\sum_{n=1}^{n_\text{max}} \frac{1}{n!} \Big(\frac{\Ub}{s}\Big)^n  e^{-nst} \frac{\mu}{n-\mu}
   \frac{\Gamma\Big(1-\frac{(n-\mu) q}{q-1}\Big)}{\Gamma\Big(1-\frac{\mu q}{q-1}\Big)} b^{-\frac{n q}{q-1}} \Bigg ]+{o}(e^{-n_\text{max} st}).
\end{equation}
Using $\langle\ln x\rangle=\lim_{\mu\to0}(1/\mu)\ln\langle x^\mu\rangle$ as before (see Eq.~\eqref{eq:xcumu}), we arrive at
\begin{equation}\label{eq:lnx-ave2}
  \langle\ln x(t)\rangle \approx \frac{q}{q-1}\ln(be^{\gamma/q}) - \sum_{n=1}^{n_\text{max}} \frac{\Gamma\Big(1+\frac{nq}{q-1}\Big)}{n\,n!} \Big(\frac{\Ub}{s}\Big)^n  b^{-\frac{n q}{q-1}} e^{-nst}+{o}(e^{-n_\text{max} st}).
\end{equation}
The first term corresponds to the result for $t\to\infty$, Eq.~\eqref{eq:lnx-ave1}, while the second term gives a correction for finite time.
This expression can be simplified further for large $q$ by using $b\approx\Ub/s$ and $\Gamma\Big(1+\frac{nq}{q-1}\Big)\approx n!$, where the latter simplification is only valid if $n_\text{max}\ln n_\text{max}\ll q$. We recognize then the expansion of $\ln$ and obtain
\begin{equation}
  \langle \ln x(t)\rangle \approx \ln(\Ub/s) + \ln\left[1-e^{-st+{1\over q}\ln{s\over\Ub}}\right]+o(e^{-n_\text{max} st}),
\label{xtfinal}
\end{equation}
where we recall that $n_\text{max}$ is such that $n_\text{max}\ln n_\text{max}\ll q$ and $n_\text{max}\ll (\Ub/s) e^{sqt}$. Furthermore, the $o(e^{-n_\text{max} st})$ is indeed a small correction only if $n_\text{max} s t \gg1$. Therefore, Eq.~\eqref{xtfinal} is only valid if $q\gg1$ (from the first condition above) and $t\Ub e^{sqt}\gg1$ (from the second and third conditions).
[We made some simplifications using $q\gg1$ to reach Eq.~\eqref{xtfinal}, but as we will see, we need to keep the term ${1\over q}\ln(s/\Ub)$ given the relevant values of $t$.]
In terms of $\tau(t)$, we finally get
\begin{equation}\label{eq:finite-time-correction}
\langle \tau(t)\rangle \approx \frac{1}{sq}\ln\left[\frac{s/\Ub}{1-e^{-st+{1\over q}\ln{s\over\Ub}}}\right]
\end{equation}
for sufficiently large $q$ and $t$.

\begin{figure}[htb]
\centerline{\includegraphics[width=3.7in,angle=270]{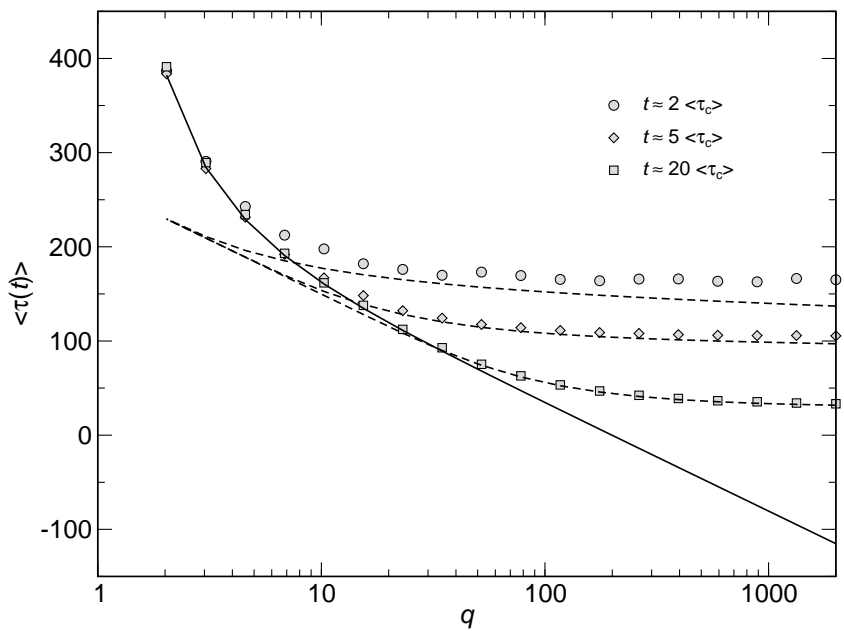}}
\caption{\label{fig:var-s_qs-fixed}Numerical evaluation of the average $\langle\tau(t)\rangle$ in simulations of the stochastic edge, as a function of $q$, for $\Ub=10^{-4}$ and $sq=0.02$ held constant throughout. Points are simulation results; the standard error from the simulations is smaller than the symbol size. As measurement times $t$, we used three multiples of $\langle\tc\rangle$, and determined $\langle\tc\rangle$ from the approximation formula Eq.~\eqref{eq:tc-approx}. The dashed lines were calculated from Eq.~\eqref{eq:finite-time-correction} (valid only for large $q$). The solid line represents $\langle\tau_{\infty}\rangle$ [Eq.~\eqref{eq:tau-ave1}, valid for all $q$].}
\end{figure}

Another way to reach Eq.~\eqref{eq:finite-time-correction} is to use the integral expression Eq.~\eqref{eq:xcumubypart} to compute $\langle\ln x(t)\rangle$. Making the change of variable $y=\lambda e^{-sqt}$, we find
\begin{equation}\label{intovery}
\langle \ln[x(t)]\rangle=-sq\langle\tau(t)\rangle=-\gamma-sqt+\int_0^{\infty}dy\ \ln(y){d\over dy}\big\langle e^{-\lambda x(t)}\big\rangle.
\end{equation}
We rewrite $\langle e^{-\lambda x(t)} \rangle$ from Eq.~\eqref{eq:momentxfinal} as a function of $y$:
\begin{equation}\label{Eq23}
\begin{aligned}
&\langle e^{-\lambda x(t)}\rangle \approx \exp\Big[-R(t) \times q y^{1-1/q}(1-y^{1/q})\Big]\qquad\text{for $y\ll1$},\\
&\text{with }R(t)={\Ub\over sq}e^{s(q-1)t}\qquad\text{and }y=\lambda e^{-sqt}.
\end{aligned}\end{equation}
[We assumed $q$ large and used $b\approx\Ub/s$.] In fact, without any approximation, one can check that $\langle e^{-\lambda x(t)}\rangle$ can be written as $\exp[-R(t)\times F(y)]$, where the function $F(y)$ has no explicit dependency on $\lambda$ or $t$. Clearly, as $y$ is proportional to $\lambda$ and $\langle e^{-\lambda x(t)}\rangle$ decreases with $\lambda$, the function $F(y)$ is an increasing function of $y$. [See for instance Eq.~\eqref{eq:momentxfinal2}, which shows how $F(y)$ increases for $y\ll1/(sq)$.] Moreover, despite the presence of the large parameter $q$, this function varies neither slowly nor rapidly with $y$, so that the speed with which $\langle e^{-\lambda x(t)} \rangle$ changes with $y$ depends only on the magnitude of $R(t)$. When $R(t)$ is large, $\langle e^{-\lambda x(t)} \rangle$ interpolates very quickly between 1 and 0, and its derivative can be approximated by a delta function. This interpolation occurs at some value $y_c$ of $y$ which is very small, hence it is justified to use Eq.~\eqref{Eq23} to compute $y_c$. Moreover, for $R(t)$ large enough, $\langle e^{-\lambda x(t)} \rangle$ becomes negligibly small within the range of validity of Eq.~\eqref{Eq23} and will go on decreasing for larger values of $y$ [because $F(y)$ is an increasing function] so that values of $y$ outside the validity range of Eq.~\eqref{Eq23} do not contribute to the integral.  All these remarks allow us to compute the integral in Eq.~\eqref{intovery}; we find, for $R(t)\gg1$,
\begin{equation}\label{Rt}\begin{aligned}
	&-sq\langle\tau(t)\rangle\approx-\gamma-sqt-\ln y_c,\\
	&R(t)\times q y_c^{1-1/q}(1-y_c^{1/q})\approx1\quad\text{with $y_c\ll1$}.
\end{aligned}\end{equation}
Eliminating $y_c$ in the previous equation gives
\begin{equation}\label{eqtauself}
\langle \tau(t)\rangle\approx{1\over s(q-1)} \ln\left[ e^{{q-1\over q}\gamma} s/\Ub\over
	1-e^{-st+s\langle\tau(t)\rangle-\gamma/q}\right].
\end{equation}
Eq.~\eqref{eqtauself} is an equation for $\langle\tau(t)\rangle$; iterating it  once and using $q$ large, we recover Eq.~\eqref{eq:finite-time-correction}, up to some negligible terms. Note that the validity condition $R(t)\gg1$ is approximatively the same as in the first method, as either can be rewritten as $t-1/(sq)\ln(sq/\Ub)\gg1/(sq)$.
 Numerical simulations (see Fig.~\ref{fig:var-s_qs-fixed}) confirm that our analytical argument is sound and that Eq.~\eqref{eq:finite-time-correction} gives indeed a good numerical approximation of the measured $\langle \tau(t) \rangle$ in stochastic edge simulations for values of $t$ larger (but not very much larger) than $\langle \tau(t)\rangle$.

We will now exploit Eq.~\eqref{eq:finite-time-correction} to test the validity of \citeauthor{DesaiFisher2007}'s result.
The purpose of computing $\langle \tau(t)\rangle$ is to compute the mean establishment time of a new class, which we call $T$ in the remainder of this section. \citet{DesaiFisher2007} take $T=\langle \tau(\infty)\rangle$. But, as we will argue now, it makes more sense to take $T\approx \langle\tau(T)\rangle$. Indeed, the reason why $\tau(t)$ depends on $t$ stems from the fact that fitting the growth of the new class by an exponential [see Eq.~\eqref{eq:deftau}] is not a perfect description of what is really hapening, and the best value of the parameter $\tau$ in this fit depends on the range of values of $t$ where we want this exponential fit to be the most precise. This range of values is precisely $t$ of order $T$, because it is at this moment that the new class becomes the second-best-fit class, starts feeding an even newer class, and becomes approximated by a deterministic exponential growth [see Eq.~\eqref{secondbestfitclass}]. The whole theory can be made self-consistent only if the value of the best-fit class just before it is established matches its value just after its establishment, which happens only if the size of the best fit class is well described for $t$ of the order of $T$. Consequently,
\begin{equation}\label{autorefT}
T\approx\langle\tau(T)\rangle.
\end{equation}
We can try to solve Eq.~\eqref{autorefT} directly from Eq.~\eqref{eq:finite-time-correction}; as the argument of the exponential in Eq.~\eqref{eq:finite-time-correction} is small for $t\approx T$, we may expand it and we obtain, ignoring subleading logarithmic terms inside the logarithm,
\begin{equation}\label{T=tau(T)}
T\approx{1\over sq}\ln{q s\over \Ub},
\end{equation}
which is quite different from the result Eq.~\eqref{tauinfty} of \citeauthor{DesaiFisher2007} and is rather closer to Eq.~\eqref{eq:rouzine}. Note that in this procedure we are operating slightly outside the range of validity of Eq.~\eqref{eq:finite-time-correction}: we need to use this equation at $t=T$ with $T$ given in Eq.~\eqref{T=tau(T)}, but we have shown it is valid only for $t-T\gg1/(sq)$, that is for $t$ slightly larger than $T$. As we use $t\approx T$ only in the argument of a large logarithm, we do not believe that this approximation should affect at all the final result Eq.~\eqref{T=tau(T)} to the leading order. Our more precise method presented in the next section of this paper confirms this claim.

\citeauthor{DesaiFisher2007}'s result  Eq.~\eqref{tauinfty} is obtained by making the approximation $T=\langle \tau(\infty)\rangle$, which is equivalent to neglecting the exponential in  Eq.~\eqref{eq:finite-time-correction}. Clearly, this procedure is only justified when it gives a result compatible to Eq.~\eqref{T=tau(T)}. This is the case only if
\begin{equation}\label{DF07valid}
  \ln q \ll \ln(s/\Ub).
\end{equation}
This finding is consistent with the arguments in \citeauthor{DesaiFisher2007}'s Appendix G. [Note that their Eq.~(G3) contains a misprint, and should use a $\gg$ sign rather than a $\ll$ sign. Michael Desai, pers.\ communication.] One way to satisfy condition~\eqref{DF07valid} is to impose $V<s$. Indeed, using Desai and Fisher's result $V=1/\langle\tau(\infty)\rangle$ with $\langle\tau(\infty)\rangle$ given by Eq.~\eqref{tauinfty}, the condition $V<s$ translates indeed into $\ln(s/\Ub)>q\gg \ln q$.

Note that in this whole section, the derivation begins by assuming that the time $T_0$ at which the second-best class starts producing mutants is $-\infty$. We shall now briefly check that this is a sound hypothesis by showing that we would have reached, to the leading order, the same final result Eq.~\eqref{T=tau(T)} by taking $T_0=0$. As can be checked from Eq.~\eqref{Eq23} and Eq.~\eqref{Rt}, the values of $\lambda$ contributing most to the integral are around $\lambda_c=y_c e^{sqt}\approx e^{sq\langle\tau(t)\rangle-\gamma}$. To reach Eq.~\eqref{T=tau(T)}, we are interested in the time $t\approx T\approx \langle\tau(T)\rangle$, for which we obtain $\lambda_c\approx qs/\Ub$, which we assumed is large. Now, if $T_0=0$, the upper bound of the integral in Eq.~\eqref{eq:momentx1} should be $\lambda$ [since we use $sq\ll1$, see Eq.~\eqref{eq:mgf-x-integral}], which is large for the relevant values of $\lambda$. As in Eq.~\eqref{simpT0}, this means we need to substract $\ln(1+1/\lambda)\approx 1/\lambda$, which is small, from the evaluation of this integral, Eq.~\eqref{integ}. But, within our working hypothesis $q\gg1$ and $\lambda e^{-sqt}\ll1$, the value of that integral is large: it diverges logarithmically for large $q$ and small $\epsilon=\lambda e^{-sqt}$; for $q>3$ and $\epsilon<.1$, it is larger than 1, which is much larger than the small correction $1/\lambda$. Therefore, considering $T_0=0$ instead of $T_0=-\infty$ does not change the final result Eq.~\eqref{T=tau(T)} at the leading order.

When using a finite back-extrapolation time, we run into another difficulty that we haven't mentioned yet. The mathematically exact value of $\langle \tau(t)\rangle$ for any finite $t$ is $-\infty$, because there is a non-zero probability $p_0$ that the size of the best-fit class is 0. For relevant values of $t$, the value of $p_0$ is incredibly small [one can show that $p_0<\exp\big[-R(t) \ln\big(1+1/(sq)\big)\big]$, with $R(t)$ given in Eq.~\eqref{Rt}]. Of course, this event never occured during all our simulations, and the only biologically observable quantity that makes sense is the average of $\tau(t)$ given that the new best-fit class is not empty. This quantity can be calculated in a precise way by replacing $\langle e^{-\lambda x(t)}\rangle$ everywhere in the previous derivation with $(\langle e^{-\lambda x(t)}\rangle-p_0)/(1-p_0)$. As a close inspection of our derivations would show, we only use the function $\langle e^{-\lambda x(t)}\rangle$ in regions where it is much larger than $p_0$, so that nothing in our final result Eq.~\eqref{eq:finite-time-correction} should be changed because of that $p_0$. We shall now however present a better, more satisfying approach where none of these issues occurs.

\bigskip

\noindent
\textbf{A better back-extrapolation:} In the previous subsection, we have seen that $\langle \tau(t)\rangle$ depends strongly on $t$. At first glance, this result is somewhat unexpected. We intended the quantity $\tau$ to be the time at which the best-fit class crosses the stochastic threshold (i.e., the establishment time of a new fitness class), and this time should have a specific, well-defined value. Instead, we have found that the expected value $\langle \tau(t)\rangle$ decays as $t$ increases, i.e., the longer we wait before we evaluate the system, the smaller the mean establishment time appears to be. This result indicates that $\tau(t)$, as defined above, is a poor method for getting an approximation of the mean establishment time.

\begin{figure}[htb]
\centerline{\includegraphics[width=3.7in,angle=270]{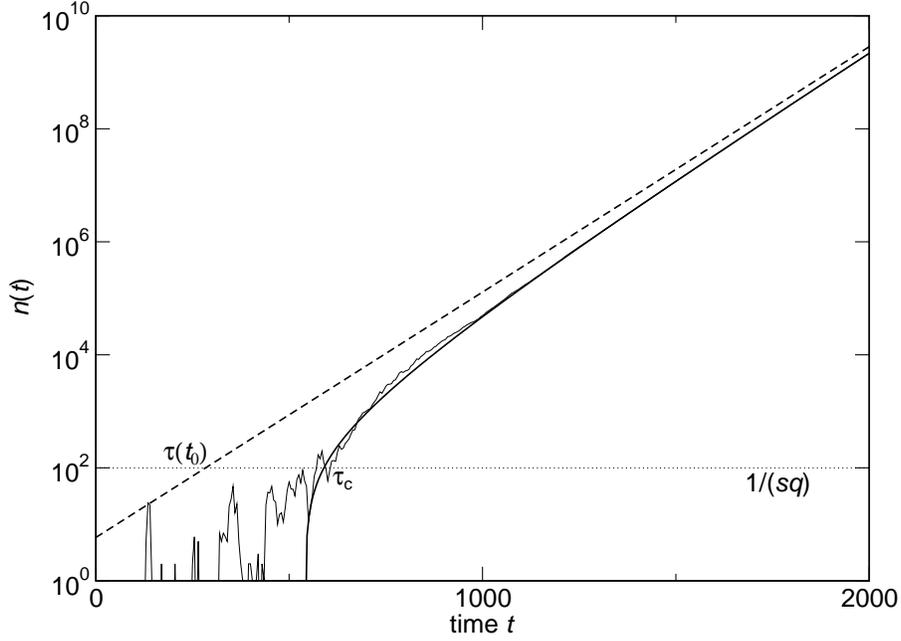}}
\caption{\label{fig:back-extrapolation}Stochastic-edge simulation and back-extrapolation to obtain $\tau(t_0)$ and $\tc$. The thin solid line represents the size $n(t)$ of the best-fit class in a typical stochastic-edge simulation run for $s=0.001$,  $\Ub=0.0001$, and $q=10$. The thick solid line is Eq.~\eqref{fitnt} with $\tc=590$ and the dashed line is Eq.~\eqref{eq:deftau} with $\tau(t_0)=284$. The values of $\tc$ and $\tau(t_0)$ have been determined at time $t_0=10000$, which means that the stochastic value $n(t_0)$ is indeed given by, respectively, Eq.~\eqref{fitnt} and Eq.~\eqref{eq:deftau}. For large times ($t\gtrsim2000$), both fits are good but for intermediate times, the stochastic $n(t)$ is best captured by the thick solid line. The time at which $n(t)$ reaches the stochastic threshold $1/(sq)$ (represented as an horizontal dotted line) is much closer to $\tc$ than to $\tau(t_0)$. Moreover, the value of $\tau(t_0)$ would have depended much more on the choice of $t_0$. For instance, taking $t_0=1000$ would have given $\tau(t_0)=380$ and $\tc=578$.}
\end{figure}

We can understand the origin of the strong time dependence of $\langle \tau(t)\rangle$ from Eq.~\eqref{eq:momentxfinal}. We rewrite this equation as
\begin{equation}
   \left\langle e^{-\lambda \left[x(t)+\frac{\Ub}{s}e^{-st}\right]}\right\rangle \approx \exp(-b\lambda^{1-1/q}).
\label{xplus}
\end{equation}
In this form, we see that the variable $x(t)$ [defined in Eq.~\eqref{defofx}] has a deterministic part $-(\Ub/s)e^{-st}$, and that the fluctuations around that deterministic part have a nearly time-independent distribution described by the generating function on the right-hand side. The deterministic part has its origin in beneficial mutations fed into the best-fit class from the second-best class, and can easily be understood by considering the deterministic approximation for the size $n(t)$ of the best-fit class:
\begin{equation}\label{eq:ode-for-n}
        \frac{dn(t)}{dt} = sqn(t) + \frac{\Ub}{sq}e^{s(q-1)t}.
\end{equation}
[This equation follows from Eq.~\eqref{deterministic} with $n_{k-1}(t)$ given by Eq.~\eqref{secondbestfitclass} and outgoing mutations neglected.] The origin of time is such that $n_{k-1}(0)=1/(sq)$, and we fix the integration constant by imposing that $n(\tc)=1/(sq)$. This choice will allow us to interpret $\tc$ later on as the establishment time, that is the time it takes to move one notch in the periodic motion of the wave. We find
\begin{equation}
n(t)=\frac1{sq}e^{sq(t-\tc)}+\frac{\Ub}{s^2q}e^{sqt}\left(e^{-s\tc}-e^{-st}\right).
\label{fitnt}
\end{equation}
Thus, because of incoming mutations, $n(t)$ does not grow purely exponentially, even in the deterministic limit. If we try to approximate this deterministic $n(t)$ or the stochastic $n(t)$ by a pure exponential as in Eq.~\eqref{eq:deftau}, the optimal fit of the parameter [$\tau$ in the case of Eq.~\eqref{eq:deftau}] depends on the time at which we want a good fit. This deviation from pure exponential growth is the source of the strong time dependence in $\langle\tau(t)\rangle$. It makes more sense to fit the stochastic $n(t)$ by Eq.~\eqref{fitnt} but with $\tc$ now a random variable (see Fig.~\ref{fig:back-extrapolation}). We may expect that in this way the distribution of $\tc$ will be largely independent of time. This is indeed the case. If we define $x(t)=sqe^{-sqt}n(t)$ as before [see Eq.~\eqref{defofx}], we obtain from Eq.~\eqref{fitnt} the deterministic evolution of $x(t)$:
\begin{equation}\label{eq:x-of-t-correct}
        x(t)+\frac{\Ub}{s}e^{-st}=\frac{\Ub}{s}e^{-s\tc}+e^{-sq\tc}.
\end{equation}
Then, comparing this equation with Eq.~\eqref{xplus}, we find that the deterministic component of $x(t)$ in the probabilistic calculation corresponds exactly to the time-dependent part of $x(t)$ in a fully deterministic model of the stochastic edge. Interpreting $\tc$ as a random variable, we see that the generating function of the right-hand side of Eq.~\eqref{eq:x-of-t-correct} is given by Eq.~\eqref{xplus} and is nearly time independent. [Only ``nearly'' because we neglected terms of order $e^{-sqt}$ in the right hand side of Eq.~\eqref{eq:momentx1} to reach Eq.~\eqref{eq:momentxfinal} and Eq.~\eqref{xplus}.]

To sum up, we write the stochastic size $n(t)$ of the best-fit class as in Eq.~\eqref{fitnt}, where $\tc$ is a random variable. We equate the mean establishment time in the full population model with $\langle\tc\rangle$. In our new approach, $\langle\tc\rangle$ does not depend much on time (the subscript ``c'' stands for constant) and we avoid the difficulty of Desai and Fisher's approach. From Eq.~\eqref{xplus} and
Eq.~\eqref{eq:x-of-t-correct} the distribution of $\tc$ is determined by
\begin{equation}\label{eq:moment-gen-K}
  \langle e^{-\lambda K}\rangle \approx \exp(-b\lambda^{1-1/q}) 
\end{equation}
with
\begin{equation}\label{Ktauc}
	 \quad K = \frac{\Ub}{s}e^{-s\tc}+e^{-sq\tc}.
\end{equation}
The new difficulty, of course, is to obtain $\langle\tau_c\rangle$ from these two equations.

\bigskip

\noindent
\textbf{Scaling function for $\langle\tc\rangle$:} The equations determining $\langle\tc\rangle$ are transcendental, and we have not been able to obtain a simple, closed-form expression for $\langle\tc\rangle$. Nevertheless, we can gain substantial insight into how $\langle\tc\rangle$ depends on the parameter values $s$, $\Ub$, and $q$. We make the change of variables
\begin{equation}\label{K'}
  K' = K\Big(\frac{\Ub}{s}\Big)^{-\frac{q}{q-1}}\quad \text{and}\quad
  X = \Big(\frac{\Ub}{s}\Big)^{-\frac{q}{q-1}} e^{-sq\tc},
\end{equation}
and obtain, from Eq.~\eqref{Ktauc} and Eq.~\eqref{K'},
\begin{align}\label{eq:tc-of-X}
  \langle \tc\rangle &= \frac{1}{s(q-1)}\ln\frac{s}{\Ub} - \frac{1}{sq}\langle\ln X\rangle,\\
  K' &= X+X^{1/q},
\end{align}
and, from Eq.~\eqref{eq:moment-gen-K},
\begin{equation}\label{eq:aveKprime}
   \langle e^{-\lambda K'}\rangle \approx \exp\Big(-c\lambda^{1-1/q}\Big)\qquad\text{with }c=\frac{bs}{\Ub}.
\end{equation}
The constant $b$ is defined in Eq.~\eqref{defb}. Inserting this definition into $c=bs/\Ub$, we obtain
\begin{equation}
  c \approx \frac{\pi}{q\sin(\pi/q)}.
\end{equation}
We note that the constant $c$ does not depend on $\Ub$ nor on $s$. Therefore, $\langle\ln X\rangle$ does not depend on $\Ub$ nor $s$, and Eq.~\eqref{eq:tc-of-X} fully captures the dependency of $\langle \tc\rangle$ both on $\Ub$ and $s$. (Actually, using the most precise version of Eq.~\eqref{defb}, there is a very weak dependency on $s$ in $c$, but for any biologically relevant case, $s$ is small and this dependency can be neglected.) We can then write
\begin{equation}\label{eq:tau-scaling}
  \langle \tc\rangle = \frac{1}{s}\Big[F(q)+\frac{1}{(q-1)}\ln\frac{s}{\Ub}\Big],
\end{equation}
where $F(q)$ is a function depending only on $q$ and given by
\begin{equation}\label{eq:basic-scaling-fct}
  F(q) = -\frac{1}{q} \langle\ln X\rangle.
\end{equation}

\begin{figure}[htb]
\centerline{\includegraphics[width=3.7in,angle=270]{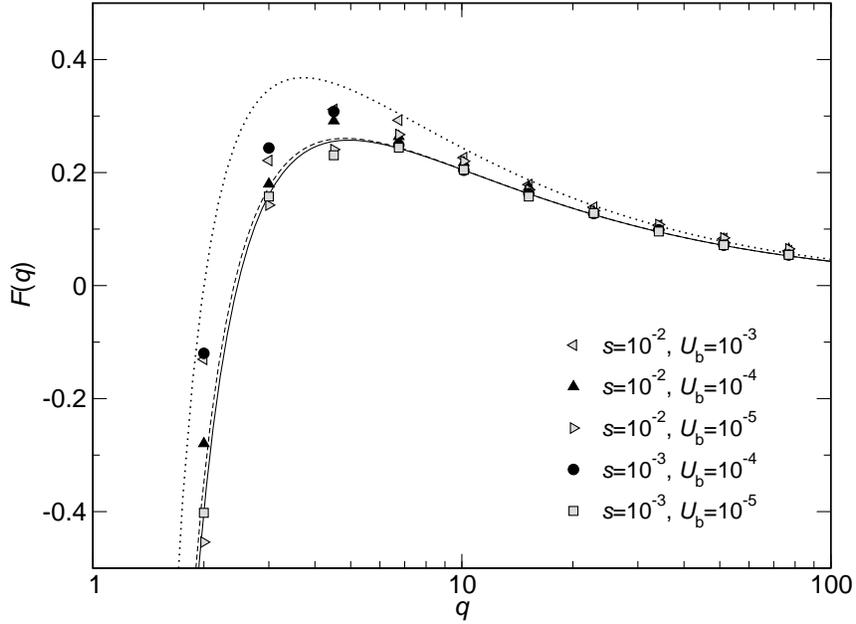}}
\caption{\label{fig:scaling-fct}Measured values $\langle\tc\rangle$ collapse onto a single scaling function $F(q)$. Data points are simulation results obtained from stochastic-edge simulations. For each parameter setting, we measured $\langle\tc\rangle$ and then plotted $s\langle\tc\rangle+\frac{1}{q-1}\ln(\Ub/s)$ as a function of $q$. The solid line represents a numerical evaluation of Eq.~\eqref{eq:basic-scaling-fct} and the dashed line represents the approximate analytic expression Eq.~\eqref{approxF}. The dotted line is the scaling function $F(q)=\frac{1}{q-1}\ln(q-1)$ derived from Eq.~\eqref{eq:tc-approx}.
}
\end{figure}

In Appendix~B, we show that we can write $\langle\ln X\rangle$ as a single integral, see Eq.~\eqref{eq:ave-lnX1}, which can be easily numerically evaluated for any value of $q$. We obtain
\begin{equation}\label{approxF}
F(q)\approx \frac{1}{q-1}[\ln(q-1) -0.345].
\end{equation}
The leading term comes from an analytical argument and the corrective term $-0.345$ is numerical. Figure~\ref{fig:scaling-fct} shows that the measured values of $\langle\tau_c\rangle$ in stocastic edge simulations can be reasonnably well collapsed on the scaling function Eq.~\eqref{approxF} for small values of $s$ and $\Ub$ in a broad interval of $q$.

Inserting Eq.~\eqref{approxF} into Eq.~\eqref{eq:tau-scaling}, we obtain
\begin{equation}\label{finaltau}
  \langle \tc\rangle \approx \frac{1}{s(q-1)}\Big[\ln\frac{s(q-1)}{\Ub}-0.345\Big].
\end{equation}
Here again, the establishment time given by Eq.~\eqref{finaltau} is very similar for large~$q$ to the result Eq.~\eqref{eq:rouzine} obtained by \citet{Rouzineetal2003,Rouzineetal2007}. 

\bigskip

\noindent
\textbf{A simple approximation formula:} With Eq.~\eqref{finaltau}, we have a good approximation for $\langle\tc\rangle$, but the derivation of this approximation was quite tedious. We can alternatively derive a simple approximation formula for $\langle\tc\rangle$ on the basis of biological considerations. A similar derivation was first presented by \citep{Desaietal2007,DesaiFisher2007}, and was also used by \citet{Rouzineetal2007} in the context of traveling wave theory.

The average total number of mutations $m(t)$ produced by the second-best class up to time $t$ is
\begin{align}
  m(t) &= \Ub\int_{-\infty}^t \frac{1}{sq}e^{s(q-1)t'}dt' \notag\\
   &= \frac{\Ub}{s^2q(q-1)}e^{s(q-1)t}.
\end{align}
Each of these mutations have a probability of going to fixation of $sq/(1+sq)\approx sq$ \citep{LenskiLevin85}. Since a single mutation that fixes is sufficient to establish a new fitness class, we have
\begin{equation}
  {sq}m(\langle\tc\rangle) \approx 1.
\end{equation}
We rearrange this equation and find
\begin{equation}\label{eq:tc-approx}
 \langle\tc\rangle \approx \frac{1}{s(q-1)}\ln\Big(\frac{s (q-1)}{\Ub}\Big).
\end{equation}
Despite the simplicity of this argument, we find that this expression has good accuracy, in particular for large $q$. Eq.~\eqref{eq:tc-approx} differs from Eq.~\eqref{finaltau} only in the constant 0.345 subtracted from the logarithm.

In the remainder of this paper, we will not use Eq.~\eqref{eq:tc-approx}. We included its derivation primarily to show that the edge treatment of \citet{Rouzineetal2007} is consistent with our derivation of $\langle\tc\rangle$.

\bigskip

\begin{figure}[htb]
\centerline{\includegraphics[width=4.5in]{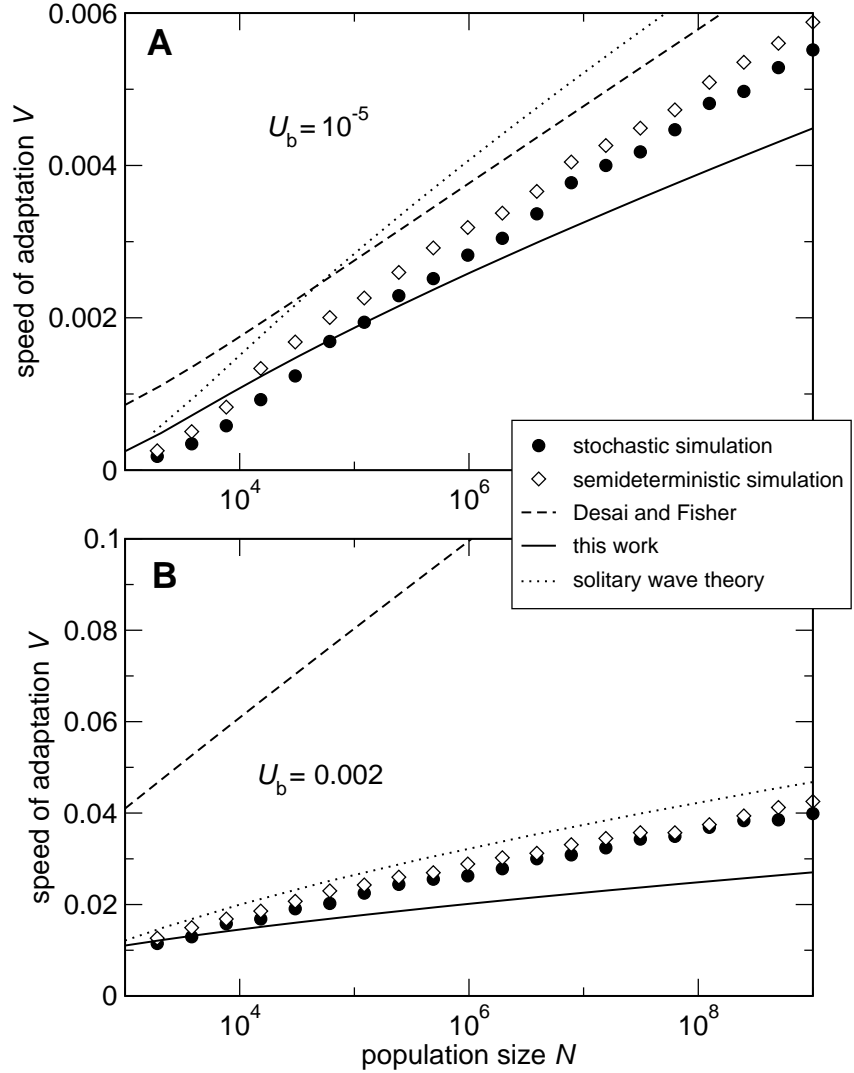}}
\caption{\label{fig:V-meas-pred}Speed of adaptation as a function of population size $N$. Points are simulation results: the solid circles come from stochastic simulations of the full model, while the open
diamonds come from semi-deterministic simulations where only the best-fit class is stochastic. Dashed lines were obtained by numerically solving Eqs.~(36) and~(39) of \citep{DesaiFisher2007}. Solid lines were obtained by numerically solving Eqs.~\eqref{finaltau} and~\eqref{eq:normalization-integral} in the present work. Dotted lines are Eq.~(52) [for part (A)] and Eq.~(51) [for part (B)] from \citep{Rouzineetal2007}. Parameters are $s=0.01$ and $\Ub=10^{-5}$ for part (A), $s=0.01$ and $\Ub=0.002$ for part (B). Note that our simulation results are in excellent agreement with simulation results reported by \citet{DesaiFisher2007}.
}
\end{figure}

\noindent
\textbf{Predicting the speed of adaptation:} The goal of calculating $\langle \tau\rangle$ in the previous subsections was to obtain the speed of adaptation $V$, which is approximately given by $1/\langle \tau\rangle$. [Throughout this subsection, we mean $\langle \tau\rangle$ to stand for either $\langle \tau(t)\rangle$ or $\langle \tc\rangle$.] Since $\langle \tau\rangle$ depends on $q$, which is a derived property of the adapting population and not known in advance, we need a second, independent expression linking $\langle \tau\rangle$ and $q$. \citet{DesaiFisher2007} obtained this second expression from the normalization condition that the sum over all fitness classes has to yield the population size $N$. They argued that at the time of establishment of the best class, the size $n_{k_0-r}$ of a fitness class $r$ mutations away from the best class is given approximately by
\begin{equation}\label{eq:bulk}
  n_{k_0-r} \approx \frac{1}{sq} \exp\Big([rq-r(r+1)/2 ] s \langle\tau\rangle\Big),
\end{equation}
because the second-best class has on average been growing exponentially at rate $s(q-1)$ for a time-interval $\langle\tau\rangle$, the third-best class has in addition been growing at rate $s(q-2)$ for an additional time-interval $\langle\tau\rangle$, and so on. As the largest term in the sum arises for $r\approx q$, \citet{DesaiFisher2007} simplified the normalization condition $N=\sum_k n_k$ to $N\approx n_{k_0-q}$, which yields
\begin{equation}\label{eq:DF-normalization}
  s\langle\tau\rangle q(q-1) \approx 2 \ln(sqN).
\end{equation}
Inserting the expression for $\langle\tau\rangle$ from \citet{DesaiFisher2007} [their Eq.~(36)] into this expression recovers their Eq.~(39), an expression that implicitly determines $q$ as a function of $s$, $\Ub$, and $N$. Note however that Eq.~\eqref{eq:DF-normalization} works only if $s\langle\tau\rangle$ is large, i.e. $s\gg V$. When $s\langle\tau\rangle$ is small, i.e. $s\ll V$, a better approximation to $N=\sum_k n_k$ is to replace the summation with an integral, $N\approx \int dk\,n_k$ which gives.
\begin{equation}\label{eq:normalization-integral}
  s\langle\tau\rangle(q-1/2)^2 \approx 2\ln(sqN)+\ln[s\langle\tau\rangle/(2\pi)].
\end{equation}
Eqs.~(\ref{eq:DF-normalization}) and (\ref{eq:normalization-integral}) correspond respectively to the two limits of a narrow wave and a broad wave discussed in \citet{Rouzineetal2007}. Here, to better compare our results to Desai and Fisher's approach, we only use Eq.~(\ref{eq:DF-normalization}) even for $s<V$ as in part~B of Fig.~\ref{fig:V-meas-pred}. Using the more correct Eq.~\eqref{eq:normalization-integral} would have resulted in only a small correction of approximately 5\% at $N=10^9$ to less than 14\% at $N=10^4$ for the parameter settings of Fig.~\ref{fig:V-meas-pred}B (data not shown).

To sum up, the final prediction in this model for the speed of adaptation $V$ is
\begin{equation}
V=\frac{1}{\langle\tc\rangle},
\end{equation}
where $\langle\tc\rangle$ as a function of $N$, $s$ and $\Ub$ is obtained by eliminating $q$ in Eq.~\eqref{finaltau} and Eq.~\eqref{eq:DF-normalization}. As we cannot analytically eliminate $q$, we have only two options: either to derive an approximate expression for $q$ from these equations, or to solve them numerically.

\citet{DesaiFisher2007} derived an approximate expression for $q$, neglecting some large logarithm inside of another logarithm [see Eqs~(39), (40) of the cited work]. The final result shown in their Fig.~5 agrees well with simulation results. However, when we compared this approximate expression to the corresponding exact numerical solution of their Eq.~(39),  we found that the term \citet{DesaiFisher2007} neglected is not small in their parameter range, and that, compared to the results of numerical simulations, the solution obtained by eliminating $q$ numerically from Eq.~\eqref{finaltau} and Eq.~\eqref{eq:DF-normalization} performed worse than their approximate expression. Thus, the performance of the approximate expression is partly due to cancellation of errors, and we will not further consider this approximate expression here.

Fig.~\ref{fig:V-meas-pred} compares how the work of \citet{DesaiFisher2007} and the present work perform in predicting the speed of adaptation $V$. The dashed lines represent the exact numerical solution to the expression derived by \citet{DesaiFisher2007}. This expression works reasonably well for low wave speeds such that $V<s$ (Fig.~\ref{fig:V-meas-pred}A), but performs poorly at high wave speeds ($V>s$, Fig.~\ref{fig:V-meas-pred}B), as expected. The poor performance at high wave speeds is caused by the breakdown of the $\langle\tau(\infty)\rangle$ approximation. If we instead use $\langle\tc\rangle$, we get a significant improvement in the prediction accuracy at high wave speeds (solid lines in Fig.~\ref{fig:V-meas-pred}). At low wave speeds, the two methods have comparable accuracies.

For comparison, we also plotted the predictions from traveling wave theory (dotted lines), as derived by \citet{Rouzineetal2003,Rouzineetal2007}. At low wave speeds (Fig.~\ref{fig:V-meas-pred}A), traveling wave theory performs approximately as well as both the original approach by \citet{DesaiFisher2007} and our revision of it. While all three methods show reasonable performance in this parameter region, none has excellent accuracy. At high wave speeds (Fig.~\ref{fig:V-meas-pred}B), traveling wave theory performs better than our revised version of the Desai-and-Fisher approach, and comes close to the speed found in semideterministic simulations (see also next paragraph). Traveling wave theory takes into account the effect of mutation pressure on intermediate fitness classes, and thus incorporates their non-exponential growth. By contrast, we have neglected this effect in the present work, and have assumed that the second-best class grows purely exponentially (Approximation~\ref{ap:exp}). Certainly, the present work tends to underestimate the speed of adaptation because of Approximation~\ref{ap:exp}. It is less clear why traveling wave theory always overestimates the wave speed. Possibly, the assumption made in traveling wave theory that the wave speed is determined by the mean size of the stochastic edge might underestimate the drag exerted by the stochastic edge when it is very small.

We also carried out semideterministic simulations in which the best-fit class was treated stochastically and all other classes were treated deterministically. The semideterministic simulation tests the fundamental assumption, made both by \citet{DesaiFisher2007} and in traveling wave theory \citep{Rouzineetal2003,Rouzineetal2007}, that only a single stochastic fitness class is necessary to describe an adapting population. Any analytical treatment of adaptive evolution based on this assumption can only ever perform as well as the semideterministic simulations. We found that the wave speed in the semideterministic simulations was close, but not exactly the same, as the true wave speed (Fig.~\ref{fig:V-meas-pred}). In general, the semideterministic simulations tended to overestimate the wave speed, in particular for small wave speeds.

\bigskip

\centerline{CONCLUSIONS}
The work by \citet{DesaiFisher2007} constitutes an interesting new approach to calculating the speed of adaptation. However, their work does not apply to high adaptation speeds, i.e., populations with large $q$. This limitation arises because the growth of the best-fit class cannot be described as a purely exponential growth times a random constant when the population evolves rapidly. Because the best-fit class is continuously being fed beneficial mutations from the second-best class, the random variable that modifies the exponential growth of the best-fit class is actually time dependent, and its mean changes with time.

Here, we have modified Desai and Fisher's method to handle correctly the non-exponential growth of the best-fit class.  Our modification leads to a substantial improvement in the prediction of the speed of adaptation for rapidly adapting populations, and agrees with predictions from traveling wave theory. However, we have relied on an exponentially growing second-best class throughout this work, even though beneficial mutations from classes with lower fitness contribute significantly to the growth of the second-best class. A more accurate treatment of adaptive evolution than we have presented here will have to take this fact into account.

\bigskip

\centerline{ACKNOWLEDGMENTS}

We would like to thank M.M.~Desai, M.~Dutour, and B.~Derrida for helpful discussions. IMR was supported by NIH grant R01 AI0639236, and COW was supported by NIH grant R01 AI065960.

\bigskip

\centerline{APPENDIX~A: VALIDITY OF APPROXIMATION \ref{ap:exp}.}

\noindent
The goal of this Appendix is to test Approximation~\ref{ap:exp}, namely that the second-best-fit class grows exponentially with a rate $s(q-1)$ :
\begin{equation}
n_{k_0-1}(t) \approx {1\over sq} e^{s(q-1)t}.
\label{secondbest}
\end{equation}
[Remember that $k_0$ was defined as the location of the stochastic edge, so that the second-best-fit class is at position $k_0-1$. The origin of time is when that class just got established: $n_{k_0-1}(0)=1/(sq)$.]

The size of any established class can be obtained from Eq.~\eqref{deterministic}, which reads for the second-best-fit class:
\begin{equation}
\frac{dn_{k_0-1}(t)}{dt}=s(q-1) n_{k_0-1}(t)+\Ub n_{k_0-2}(t) -\Ub n_{k_0-1}(t).
\label{det2}
\end{equation}
Eq.~\eqref{secondbest} is the solution of Eq.\eqref{det2} only if the second and third terms on the right-hand side of Eq.\eqref{det2} are negligible. The third term is easily dealt with as we assumed throughout this work that $s(q-1)\approx sq \gg \Ub$. For the second term, we need to evaluate the size $n_{k_0-2}(t)$ of the third-best class.

We shall proceed by assuming that the third-best class  is described by a deterministic exponential formula, analogous to the equation used in the two-class model for the second-best class. Then, from Eq.~\eqref{eq:bulk}, we obtain for $0\le t\le\tau$ the expression
\begin{equation}
n_{k_0-2}(t)={1\over sq} e^{s(q-1)\tau+s(q-2)t}.
\end{equation}
This expression is based on the assumptions that the class $k_0-2$ got established at time $-\tau$ when it reached the size $1/(sq)$, grew from time $-\tau$ to time 0 with rate $s(q-1)$, and grows from time 0 to time $\tau$ with rate $s(q-2)$.

Using the value of $\tau$ given in Eq.~\eqref{finaltau}, we find
\begin{equation}
n_{k_0-2}(t)=\frac{\alpha}{\Ub}e^{s(q-2)t},
\end{equation}
where $\alpha$ is of order 1. Using $n_{k_0-1}(0)=1/(sq)$ and neglecting the third term on the right-hand side of Eq.~\eqref{det2}, we find as the solution to Eq.~\eqref{det2}
\begin{equation}
\label{3dclassfinal}
n_{k_0-1}(t)= \frac{1}{sq}e^{s(q-1)t}+\frac{\alpha}{s}\left(e^{s(q-1)t}-e^{s(q-2)t}\right).
\end{equation}
For moderate times such that $t\ll1/(sq)$, the second term in Eq.~\eqref{3dclassfinal} is negligible and we recover Eq.~\eqref{secondbest}. However, the most relevant time interval is when $t$ is very close to $\tau$, when most of the mutations from the second-best class to the not-yet-established best class occur \citep{Rouzineetal2007}. For $t\sim \tau$, further estimates depend on whether product $s\tau$ is small or large (i.e., whether $V \gg s$ or $V\ll s$). For $s\tau\ll1$, using $\exp[s(q-2)t]
\approx(1-st)\exp[s(q-1)t]$, we obtain

\begin{equation}
\label{secondbest2}
n_{k_0-1}(t)\approx \left(\frac{1}{sq}+\alpha t\right)e^{s(q-1)t}.
\end{equation}
This expression deviates from  Eq.~\eqref{secondbest} due to the second
term in parentheses. The deviation is by a factor of order 2 when $t$ becomes of the order of $1/(sq)$, which happens early in a cycle, as $\tau\gg1/(sq)$ from Eq.~\eqref{finaltau}. At the end of cycle, $t\sim \tau$, the second term is larger than the first term by a factor of $\ln(sq/U_b)\gg 1$. Therefore, Approximation~\ref{ap:exp} is not valid, and the second-best-fit class cannot be described by Eq.~\eqref{secondbest}. 

At $s\tau \gg 1$ and $t\sim \tau$, we can neglect the third exponential in Eq.~\eqref{3dclassfinal}. Then, instead of Eq.~\eqref{secondbest2}, we obtain
\begin{equation}
\label{secondbest3}
n_{k_0-1}(t)\approx \left(\frac{1}{sq}+\frac{\alpha}{s}\right)e^{s(q-1)t}.
\end{equation}
The first term in parenthesis is negligible and the result differs from Eq.~\eqref{secondbest}  by the large factor $\alpha q\gg 1$. Therefore, Approximation~\ref{ap:exp} is not valid in this case either.

Thus, taking into account the third-best class creates an additional large factor in the size of the second-best class at the most relevant times $t\sim \tau$. This factor is on the order of either $q$ or $\ln(sq/U_b)$, whichever is smaller, and approximation~\ref{ap:exp} is not valid by itself. One could try to fix this issue by using Eq.~\eqref{3dclassfinal} instead of Eq.~\eqref{secondbest} for the size of the second best class, but it would make the derivation much more complicated. Note however that, as the effects of mutations only enter the final result through the logarithm of the mutation rate, it is plausible (but remains to be checked) that the large corrective factors of Eq.~\eqref{secondbest2} or Eq.~\eqref{secondbest3} will enter the final result as a logarithmic correction. On the other hand, there is no guarantee that taking into account the third-best class is sufficient, and it might be that one needs also to consider the effects of the fourth or fifth-best class. In all cases, the replacement of the full population model by a two-class model with an exponentially growing second-best class is problematic and deserves a more careful investigation.

\bigskip

\centerline{APPENDIX~B: CALCULATING $\langle \ln X\rangle$}

\noindent
In order to calculate $F(q)$, we have to calculate $\langle \ln X\rangle$, where $X$ is the
only positive root of
\begin{equation}
 X+X^{1/q} =K
\label{defX}
\end{equation}
and we have written $K$ instead of $K'$ for simplicity. The moment generating function for $K$ is [Eq.~\eqref{eq:aveKprime}]:
\begin{equation}\label{elK}
\langle e^{-\lambda K}\rangle=\exp[-c\lambda^{1-1/q}].
\end{equation}

A first approach is to evaluate $\langle\ln X\rangle$ numerically using an inverse Laplace transform. First, we calculate the density function $p_{K}(y)$ of the probability distribution of $K$ from the inverse Laplace transform ${\cal L}^{-1}$ of the moment generating function of $K$:
\begin{equation}
  p_{K}(y) = {\cal L}^{-1}\{\exp(-c\lambda^{1-1/q})\},
\end{equation}
where ${\cal L}^{-1}$ can be written as an integral. In practice, this integral can be evaluated with efficient numerical algorithms \citep{ValkoAbate2004,AbateValko2004}. A transformation of variables gives us the density function $p_{X}(y)$ of the probability distribution of $X$:
\begin{equation}
  p_X(y) = \Big(1+\frac{1}{q}y^{(1/q)-1}\Big)p_{K}(y+y^{1/q})
\end{equation}
Finally, we integrate to obtain $\langle \ln X\rangle$:
\begin{equation}\label{eq:ave-lnX2}
  \langle \ln X\rangle = \int_0^\infty p_X(y) \ln y\, dy.
\end{equation}
This method can be worked out, but it is delicate and time expensive to evaluate numerically with a good accuracy these not so well behaved double integrals, especially for large values of $q$. We now present an alternative method which allows us to write $\langle \ln X\rangle$ as a simple integral, which is much easier to evaluate.

\bigskip

\noindent
\textbf{Writing $\ln X$ as a series:} Our first step is to invert Eq.~\eqref{defX}. By using Cauchy's integral formula from complex analysis, we can write for any analytical function $f$ the quantity $f(X)$ as
\begin{equation}
f(X)=\frac{1}{2\pi i}\oint f(z) \frac{1+\frac{1}{q}z^{(1/q)-1}}{z+z^{1/q}-K}dz,
\end{equation}
where the integration is on a contour surrounding the only positive root
of Eq.~\eqref{defX}. We set $f(z)=z^\mu$, and make use of the Taylor-series
\begin{equation}
  \frac{1}{a+b-K} =\sum_{n\ge0}(-1)^n \frac{b^n}{(a-K)^{n+1}}.
\end{equation}
Setting $a=z$, $b=z^{1/q}$ in the above expansion,
we obtain
\begin{align}
X^\mu&=\frac{1}{2\pi i}\oint z^\mu \sum_{n\ge0}(-1)^n \frac{\left(1+\frac{1}{q}z^{(1/q)-1}\right)z^{n/q}}{(z-K)^{n+1}}dz\\
&=\sum_{n\ge0}(-1)^n \frac{1}{2\pi i}\oint \frac{(z+K)^{\mu+n/q}+  \frac{1}{q}
(z+K)^{\mu+n/q+(1/q)-1}}{z^{n+1}}dz\\
&=\sum_{n\ge0}(-1)^n \left[\binom{\mu+n/q}{n} K^{\mu+n/q-n}+ \frac{1}{q}
\binom{\mu+n/q+(1/q)-1}{n} K^{\mu+n/q+(1/q)-1-n}\right]\\
&=K^\mu+\sum_{n\ge1}(-1)^n K^{\mu+n/q-n} \frac{q\mu}{q\mu+n}\binom{\mu+n/q}{n}.
\end{align}
[We use the Binomial symbol $\binom{x}{n}$ for $x$ non-integer, with the convention that $\binom{x}{n}=x(x-1)\cdots(x-n+1)/n!\,$.] Expanding both sides of the equation to first order in $\mu$ and comparing the coefficients of the linear term, we find
\begin{equation}\label{eq:X-of-K}
\ln X =\ln K +\sum_{n\ge1}(-1)^n \frac{q}{n}\binom{n/q}{n}K^{n/q-n}.
\end{equation}

\bigskip

\noindent
\textbf{Taking the average:} We can now calculate $\langle \ln X\rangle$ by averaging Eq.~\eqref{eq:X-of-K} term by term. Following the same steps as in the derivation of Eqs.~\eqref{eq:avexpowmu} and~\eqref{eq:lnx-ave1} in the main text, we obtain from Eq.~\eqref{elK}
\begin{align}\label{eq:Kpowmuave}
\langle K^\mu\rangle & =\frac{\Gamma\big(1-\frac{\mu q}{q-1}\big)}{\Gamma(1-\mu)}c^{\frac{\mu q}{q-1}}\qquad \text{for $\mu<0$,}\\
\label{eq:lnKave}
\langle\ln K\rangle &=\frac{q}{q-1}\ln\big(c e^{\gamma/q}\big).
\end{align}
Using these two equations, we find
\begin{align}
\langle\ln X\rangle&=\frac{q}{q-1}\ln\left( c
e^{\gamma/q}\right)+\sum_{n\ge1}(-1)^n \frac{q}{n}
\binom{n/q}{n}\frac{\Gamma(1+n)}{\Gamma(1+n-n/q)}c^{-n}\\
&=\frac{q}{q-1}\ln\left( c e^{\gamma/q}\right)-\sum_{n\ge1}\frac{q^2}{\pi
n^2(q-1)}\Gamma(1+n/q)\sin(\pi n/q)c^{-n},
\end{align}
where we have made use of Euler's reflection formula $\Gamma(x)\Gamma(1-x)=\pi/\sin(\pi x)$. The resulting series diverges. However, we will treat it as a formal expansion of $\langle\ln X\rangle$ and continue. We
replace $\Gamma(1+n/q)$ by its integral representation, integrate
by parts once, and obtain
\begin{equation}
\langle\ln X\rangle=\frac{q}{q-1}\left[\ln\left( c e^{\gamma/q}\right)-
\int_0^\infty d\lambda\ \frac{e^{-\lambda}}{\pi\lambda}\sum_{n\ge1}\frac{1}{n} \lambda^{n/q}\sin(\pi
n/q)c^{-n}\right].
\end{equation}
We now write $\sin(\pi n/q)$ as the imaginary part of $e^{i\pi n/q}$, and notice that the remaining sum is the Taylor expansion of the complex logarithm. Thus, we arrive at
\begin{equation}
\langle\ln X\rangle=\frac{q}{q-1}\left[\ln\left( c e^{\gamma/q}\right)-\int_0^\infty d\lambda\ \frac{e^{-\lambda}}{\lambda\pi}
\Im\left(-\ln\left[1-\frac{\lambda^{1/q}e^{i\pi/q}}{c}\right]\right)\right],
\end{equation}
where $\Im(z)$ indicates the imaginary part of $z$. Let $\rho>0$ and $\phi$ be such that
$\rho e^{-i\phi}= 1-\lambda^{1/q}e^{i\pi/q}/c$. Then, we have
\begin{equation}
\tan\phi =
\frac{\sin(\pi/q)\lambda^{1/q}/c}{1-\cos(\pi/q)\lambda^{1/q}/c}\qquad\text{with $\sin\phi\ge0$}
\end{equation}
and
\begin{equation}
\Im\left(-\ln\left[1-\frac{\lambda^{1/q}e^{i\pi/q}}{c}\right]\right)=\phi.
\end{equation}
The angle $\phi$ is defined only up to a multiple of $2\pi$, but the
result must be $\phi=0$ when $\lambda=0$ and the result must be a
continuous function of $\lambda$. (While the sum converges only when $\lambda^{1/q}/c < 1$, we are considering here the analytical continuation of this function.) This reasoning implies that $0\le\phi\le\pi$.

The final result is then
\begin{equation}\label{eq:ave-lnX1}
\langle\ln X\rangle
=\frac{q}{q-1}\left[\ln\left( c e^{\gamma/q}\right)-\int_0^\infty d\lambda
\frac{e^{-\lambda}}{\lambda\pi}\phi(\lambda)\right],
\end{equation}
where
\begin{equation}\label{eq:phi-condition}
\tan\phi(\lambda)=\frac{\sin(\pi/q)}{c \lambda^{-1/q}-\cos(\pi/q)}
\quad\text{and}\
0\le\phi(\lambda)\le\pi.
\end{equation}
Note that
\begin{equation}
\phi(\lambda)=\begin{cases}
\displaystyle\arctan\left(\frac{\sin(\pi/q)}{c
\lambda^{-1/q}-\cos(\pi/q)}\right)&\text{when
$c\lambda^{-1/q}>\cos(\pi/q)$,}\\[2ex]
\displaystyle\pi+\arctan\left(\frac{\sin(\pi/q)}{c \lambda^{-1/q}-\cos(\pi/q)}\right)&\text{when
$c\lambda^{-1/q}<\cos(\pi/q)$.}
\end{cases}\end{equation}

We evaluated both Eq.~\eqref{eq:ave-lnX1} and Eq.~\eqref{eq:ave-lnX2} numerically, and found excellent agreement between the two formulas.

\bigskip\noindent\textbf{Leading asymptotic of $\langle\ln X\rangle$:}
We now evaluate the integral in Eq.~\eqref{eq:ave-lnX1} in the large $q$ limit. For a fixed small $\lambda$ and $q\to\infty$, it is easy to see that 
\begin{equation}
\phi(\lambda)\approx -\frac{\pi}{\ln\lambda}\qquad\text{for fixed (small) $\lambda$ and $q\to\infty$.}
\label{phiapp}
\end{equation}
(Remember that $c\approx1$ for large $q$.) However, replacing $\phi(\lambda)$ by that expression leads to a diverging integral. What happens is that for a given large $q$, the approximation Eq.~\eqref{phiapp} breaks for extremely small values of $\lambda$, and we obtain
\begin{equation}
\phi(\lambda)\approx \frac{\pi\lambda^{1/q}}{q}\qquad\text{for fixed (large) $q$ and $\lambda\to0$.}
\label{phiapp2}
\end{equation}
With the latter approximation, the integral converges. Looking more closely at the approximations made, we can check that Eq.~\eqref{phiapp} is valid for $e^{-q}\ll\lambda\ll 1$ and that Eq.~\eqref{phiapp2} is valid for $\lambda\ll e^{-q}$.

Therefore, it makes sense to cut the integral into three parts. One for $0<\lambda<e^{-q}$, where we use Eq.~\eqref{phiapp2}, one for $e^{-q}<\lambda<\epsilon$, where we use Eq.~\eqref{phiapp} and where $\epsilon$ is some fixed small number, and one for $\lambda>\epsilon$. It is easy to check that the first and third parts give a number of order 1 (in other words, they do not diverge when $q\to\infty$) and that the second part dominates the integral:
\begin{equation}
\int_0^\infty d\lambda
\frac{e^{-\lambda}}{\lambda\pi}\phi(\lambda)\approx\int_{e^{-q}}^\epsilon d\lambda\frac{-1}{\lambda\ln\lambda}=\ln\big[-\ln(e^{-q})\big]-\ln\big[-\ln\epsilon\big]\approx\ln q.
\end{equation}
Once this leading term has been identified, we can evaluate numerically the integral for many values of $q$ and extract an asymptotic expansion of the correction to the leading term. We found:
\begin{equation}
\int_0^\infty d\lambda
\frac{e^{-\lambda}}{\lambda\pi}\phi(\lambda)\approx
\ln q - 0.345 -\frac{0.45}{q}+\frac{1.0}{q^2}-\frac{0.3}{q^3}+\cdots.
\end{equation}
Another possibility is to write an expansion of the integral in powers of the variable $q-1$:
\begin{equation}
 \int_0^\infty d\lambda
  \frac{e^{-\lambda}}{\lambda\pi}\phi(\lambda)  \approx
   \ln(q-1) -0.345 + \frac{0.58}{q-1} + \frac{0.19}{(q-1)^2} +\cdots.
\label{numexpn}
\end{equation}
Both asymptotic expansion are, of course, equally good for large $q$, but it happens that truncated to its first terms, the second expansion is better than the first at approximating the integral for smaller $q$. Inserting the latter expansion into Eq.~\eqref{eq:ave-lnX1} and using Eq.~\eqref{eq:basic-scaling-fct}, we recover Eq.~\eqref{approxF}.

We have not been able to find a theory for the numerical coefficients of this asymptotic expansion, and this remains an interesting challenge. The expansion of Eq.~\eqref{numexpn} is a very good approximation of the integral in the range $q\in[2,\infty)$.

\cleardoublepage

\section*{Figures}

\end{document}